\documentclass[prd,aps,showpacs,floats]{revtex4}
\usepackage{amsmath,amssymb}
\usepackage{graphicx}
\usepackage{epsf}

\renewcommand{\baselinestretch}{1.0}

\newcommand{\be}{\begin{equation}}
\newcommand{\ee}{\end{equation}}
\begin{document}
\topmargin 0pt
\oddsidemargin=-0.4truecm
\evensidemargin=-0.4truecm
\renewcommand{\thefootnote}{\fnsymbol{footnote}}
\overfullrule 0pt
\date{\today}

\title{\
\vglue -2.0cm
{\small \hfill hep-ph/0309312}\\
\vglue 1.0cm
Atmospheric neutrinos: LMA oscillations, 
$U_{e3}$ induced interference and CP-violation}
\author{O. L. G. Peres$^{1,2}$}\email{orlando@ifi.unicamp.br} 
\author{A. Yu. Smirnov$^{2,3}$}\email{smirnov@ictp.trieste.it}

\affiliation{\\ \\
 $^1$ Instituto de F\'{\i}sica Gleb Wataghin,
 Universidade Estadual de Campinas -- UNICAMP, 13083-970 Campinas,
 Brazil \\
 $^2$ The Abdus Salam International Centre for Theoretical Physics,  
I-34100 Trieste, Italy \\
 $^3$ Institute for Nuclear Research of Russian Academy 
of Sciences, Moscow 117312, Russia}

\begin{abstract}

We consider oscillations of the low energy (sub-GeV sample) 
atmospheric neutrinos in the three neutrino context. 
We present the semi-analytic study of  the neutrino evolution  and 
calculate  characteristics  of  the $e$-like events (total number, 
energy spectra and  zenith angle distributions) 
in the presence of  oscillations. At low energies there are three 
different contributions to the number of events: the LMA contribution (from  
$\nu_e$-oscillations driven by the solar oscillation parameters), 
the $U_{e3}$-contribution proportional to $s_{13}^2$, and 
the $U_{e3}$ - induced interference of the two amplitudes driven 
by the solar oscillation parameters. 
The interference term is sensitive to the CP-violation phase. 
We describe in details properties of these contributions. 
We find that the LMA, the interference and 
$U_{e3}$  contributions  can reach 5 - 6\% , 2 - 3\% and 
1 - 2 \% correspondingly.  
An existence of the significant ($ > 3 - 5\%$) excess of the $e$-like
events in the sub-GeV sample  
and the absence of the excess in the multi-GeV range  testifies for 
deviation of the 2-3 mixing from maximum. We consider a possibility to 
measure the deviation as well as the CP- violation phase in future
atmospheric neutrino  studies.

\end{abstract}


\maketitle

\renewcommand{\thefootnote}{\arabic{footnote}}
\setcounter{footnote}{0}
\renewcommand{\baselinestretch}{0.9}
\section{Introduction}

The existing results on atmospheric neutrinos~\cite{SuperK} 
are well described in terms of pure 
$\nu_\mu\leftrightarrow \nu_\tau$ oscillations with maximal 
or close to  maximal mixing. 
Analysis of the Super-Kamiokande data gives~\cite{SuperK,sk-eps2003}
mass squared difference and mixing in the interval:  
\begin{equation}
\Delta m^2_{32} = (1.3 - 3.0) \times 10^{-3} {\rm eV}^2~, ~~~
\sin^2 2\theta_{23} > 0.9,   ~~~~ (90 \% ~ {\rm  C.L.}) . 
\label{atmdata}
\end{equation}
The results of SOUDAN \cite{SOUDAN} and MACRO \cite{MACRO} experiments are  in a good  
agreement with (\ref{atmdata}). 
The oscillation interpretation (\ref{atmdata}) has been  further confirmed by the results of 
the K2K experiment  \cite{K2K}.

Till now no compelling evidence of oscillations of the 
atmospheric $\nu_e$ has been obtained. 
The  $3\nu$ global analysis of the atmospheric neutrino data 
is in agreement with no $\nu_e$-oscillations~\cite{glob3}. The best fit point coincides  
with zero $\sin \theta_{13}$ within $1\sigma$ \cite{glob3}. 

At  the same time, after the first KamLAND result~\cite{Kam} we can definitely 
say that the $\nu_e$-oscillations of  atmospheric neutrinos should 
appear at some level. Indeed, KamLAND has confirmed the large mixing MSW 
(LMA-MSW) solution of the solar neutrino problem. The combined analysis 
of the solar and KamLAND data leads to  values of the oscillation 
parameters~\cite{comb}: 
\be
\Delta m^2_{21} = (5 - 10)\cdot 10^{-5}~ {\rm eV}^2~,~~
\tan^2 \theta_{12} =  0.3 - 0.5~.
\label{solar}
\ee
The parameters (\ref{solar}), which we will call the LMA parameters,  
should lead to oscillations of the $\nu_e$  
component in the atmospheric neutrino flux. 

The oscillations driven by the LMA  parameters have been discussed before
~\cite{yasuda,thun,Fogli,kim,sakai99,brunn,orl1,Gonzalez-Garcia:2002mu}. 
It was marked in Ref.~\cite{Fogli} that the effect 
of sub-leading oscillations driven by  $\Delta m^2_{21}$ 
is significant only for the sub-GeV events and the size of effect is  
at the level of the statistical errors. In Ref.~\cite{kim} it was
argued that the excess of e-like events in the sub-GeV sample favors
the large mixing MSW solution of the solar neutrino problem. 

The detailed study of the effect has been performed in  our previous  paper
\cite{orl1},  where  it was  shown that the neutrino oscillations with  LMA parameters 
can  lead to an observable (up to 10 - 12 \%) excess of the 
$e$-like events in the sub-GeV atmospheric neutrino sample. 
The excess has a weak zenith angle dependence 
in the low energy part of the sample and a strong zenith 
angle dependence in the high energy part. 
The excess rapidly decreases with energy of neutrinos, and it 
is strongly suppressed in the multi-GeV range. These 
signatures allow one to disentangle the effect of oscillations 
due to solar $\Delta m^2$  from other possible explanations of the excess.  

It was shown that the relative excess is determined by the 
two neutrino transition probability $P_2$ and  the ``screening" factor:   
\be
\frac{F_e}{F_e^0} - 1 =  P_2(r \cos^2 \theta_{23} - 1)~,
\label{fluxe}
\ee 
where $F_e$ and $F_e^0$ are the electron neutrino fluxes with and 
without oscillations,  $r$ is the ratio of the original muon and
electron neutrino fluxes.  
The screening factor (in brackets)  is related to existence of both
the electron and muon neutrino  in the original atmospheric neutrino flux. 
The appearance of excess (or deficiency) depends strongly on deviation of the 
$\nu_{\mu} - \nu_{\tau}$ (or 2 - 3) mixing responsible for 
the dominant mode of the atmospheric  neutrino oscillations 
from maximal value.  Indeed, in the sub-GeV region $r \approx 2$,  so 
that the screening factor is very small 
when the $\nu_{\mu} - \nu_{\tau}$  
mixing is maximal. 
Due to this factor the excess is in general small even though  
the $2\nu-$ probability can be of the order 1.  
The probability $P_2$,  and consequently the excess,   increase rapidly with  
$\Delta m^2_{21}$.

As far as  the experimental results are concerned,  there is a  hint that 
some  excess of the $e-$like events indeed exists in the sub-GeV range. Furthermore, 
the excess increases with decrease of energy within the sample~\cite{super-data-used}.  
In comparison with predictions  based on  
the atmospheric neutrino flux from ref.\cite{honda} the excess 
is about (12 - 15)\% in the low energy part of the sub-GeV sample 
($p < 0.4$ GeV, where $p$ is the momentum of lepton).  
It  has no significant zenith angle  dependence.   
In the higher energy part of the sub-GeV sample ($p >  0.4$ GeV) 
the excess is about 5\% , and practically there is no excess 
in the multi-GeV region ($p > 1.33$ GeV).

The excess is within estimated  20\% uncertainty in the original 
atmospheric neutrino flux. The analysis of  data with 
free overall normalization leads to the best fit 
$e-$like signal which practically excludes the
excess~\cite{super-data-used,totsuka,sk-eps2003}. 
However, the recent data on  primary cosmic rays~\cite{BESS,AMS} as well as new
calculations  of the atmospheric neutrino
fluxes~\cite{3d-nufluxes} change the  situation. New results imply lower
neutrino flux, and therefore larger   excess which is difficult to
explain by change of normalization~\cite{super-data-used,totsuka}.  

The $\nu_e$  oscillations  can be also induced by non-zero 1-3 mixing 
and $\Delta m^2_{31}$ responsible for the dominant mode of the atmospheric neutrino 
oscillations~\cite{yasuda,threenu1,threenu2,threenu3,threenu4,four1,ADLS,Bernabeu,Gonzalez-Garcia:2002mu,Gonzalez-Garcia:2003qf,Maltoni:2003da}. 
These oscillations require non-zero value of mixing matrix element $U_{e3}$. 
They are reduced to the vacuum oscillations for the sub-GeV sample. 
For the multi-GeV sample the Earth matter effect becomes important
which can enhance  the oscillations~\cite{ADLS}.
For neutrinos  which cross the core,   the dominant effect is the parametric
enhancement of oscillations. The size of effect is 
restricted by  the  CHOOZ bound on  $U_{e3}$~\cite{CHOOZ}. 

In this paper we will further study the oscillation effects driven 
by the LMA oscillation parameters using an updated experimental information. 
We will consider an additional effect in the sub-GeV 
sample  induced by non-zero 1-3 mixing.  
We study  effects of the interplay of 
oscillations with the LMA parameters and non-zero $U_{e3}$. 
In particular, we will discuss the interference induced by non-zero 
$\sin \theta_{13}$.  
Some preliminary results of this study have been published in \cite{PS01}. 
The interference term depends on the CP-violation phase. We calculate 
effects of CP-violating phase and estimate a  possibility to observe it 
in future atmospheric neutrino experiments. 

The paper is organized as follows. In Sec. II we present the semi-analytical 
study of  evolution of  $3\nu$ system at low neutrino energies 
which correspond to the sub-GeV 
sample of events. We present the $3\nu$ neutrino transition probabilities in terms 
of the two neutrino probabilities. We calculate the latter numerically and 
study their  properties. In Sec. II.C we give general expression 
for the flux of  electron neutrinos. 
In Sec. III. we calculate the number of $e$-like events in the water 
Cherenkov detectors with and without oscillations. 
We consider the zenith angle and  energy distributions of these events. 
We study separately effects of the LMA-oscillations (Sec. III.A), the 
$U_{e3}$ induced interference (Sec. III.B) and the  CP-violation 
(Sec. III.C).  In Sec. IV we discuss a possibility to measure the 
deviation of  2 - 3 mixing from maximum as well as the CP-violating phase in 
future atmospheric neutrino experiments. 
Discussion of the  results and conclusions are given in sect. V.

\section{Evolution of the neutrino system}

We consider the three-flavor neutrino system  with  
hierarchical mass squared differences: 
$\Delta m^2_{21} = \Delta m^2_{\odot}<< \Delta m^2_{31} = \Delta
m^2_{atm}$ (see Eqs.~(\ref{atmdata},\ref{solar})). 
The evolution of the neutrino vector, $\nu_f \equiv (\nu_e,
\nu_{\mu}, \nu_{\tau})^T$,  is described by the equation 
\be
i \frac{d \nu_f}{dt} =
\left( \frac{U M^2 U^\dagger}{2 E} + \hat V \right) \nu_f, 
\label{evolution}
\ee 
where $E$ is the neutrino energy,   
$M^2 \equiv  diag(0, \Delta m_{21}^2, \Delta m_{31}^2)$
is the diagonal matrix of neutrino mass squared eigenvalues,  
$\hat V \equiv diag(V, 0 ,0)$ is the matrix of matter-induced neutrino potentials
with $V = \sqrt 2 G_F N_e$, $G_F$ and $N_e$ being the Fermi constant 
and the electron number density  respectively.   
The mixing matrix $U$ is defined through $\nu_f = U \nu_{mass}$, where
$\nu_{mass} \equiv 
(\nu_1, \nu_2, \nu_3)^T$ is the vector of neutrino mass
eigenstates. The matrix can  be parameterized as 
\be
U = U_{23} D_{\delta_{CP}} U_{13} U_{12}, ~~~D_{\delta_{CP}} \equiv
diag (1, 1, e^{i\delta_{CP}}),    
\ee
where $U_{ij}= U_{ij}(\theta_{ij})$ performs the rotation  
in the $ij$- plane on the angle $\theta_{ij}$. This parameterization coincides with the 
standard one up to (unphysical) renormalization of the mass eigenstates: 
$\nu_{mass} = D_{\delta_{CP}}^* \nu_{mass}'$.

\subsection{Propagation basis}

The dynamics of oscillations is simplified in the ``propagation" basis 
$\tilde{\nu} = (\tilde{\nu_e}, \tilde{\nu}_{2}, \tilde{\nu}_{3})^T$,   
which is related to  the flavor basis by the  unitary transformation  
\be 
\nu_f = \tilde{U} \tilde{\nu}, 
\label{p-basis}
\ee
and the matrix $\tilde{U}$  can be introduced in the following way. First,
let us  perform the  rotation  $\nu_f =  U_{23} D_{\delta_{CP}}  U_{13} \nu'$. Using
Eq.~(\ref{evolution}) we find that  the instantaneous Hamiltonian for
the new states $\nu'$ takes the form 

$$
H' =  \frac{1}{2E} U_{12} M^2
U^\dagger_{12} +  U^\dagger_{13} V  U_{13}~, 
$$ 
or explicitly 
\be
{H'} =
\left(\begin{array}{ccc}
s_{12}^2 \Delta m_{21}^2/2E + V c_{13}^2   &  s_{12}c_{12}\Delta
m_{21}^2/2E  &  V s_{13} c_{13}   \\
s_{12}c_{12} \Delta m_{21}^2/2E & c_{12}^2 \Delta m_{21}^2/2E  &   0  \\
 V_e s_{13} c_{13} &  0  & \Delta m_{31}^2/2E  + V s_{13}^2
\end{array}\right)\, , 
\label{matr1}
\ee
($c_{12} \equiv \cos \theta_{12}$, $s_{12} \equiv \sin \theta_{12}$,  etc.). 
Since the transformation  $U_{23} D_{\delta_{CP}} U_{13}$ is constant (no
dependence on density and therefore time),  
the evolution equation for  $\nu'$ is given by the Schr\"odinger
equation with the Hamiltonian 
$H'$.  Notice that the CP-violation phase is removed from the equation
for $\nu'$, and it appears in the  projection of the 
flavor basis on  $\nu'$ only. So, the evolution of system is CP-symmetric. 

Second,  let us   make an additional $(\nu_e' - \nu_{\tau}')$ rotation,  
$\nu' =  U_{13}' \tilde{\nu}$, which removes the off-diagonal terms
$H_{13}', H_{31}'$ of the Hamiltonian (\ref{matr1}). 
The angle of this rotation depends on the density: 
\be
\tan (\Delta \theta_{13}) \approx s_{13} c_{13} \frac{2EV}{\Delta
m^2_{13}}.
\label{angle-m}
\ee
Since the low energy part of the sub-GeV  sample is produced by  neutrinos
with  energies (0.1 - 0.5) GeV the factor in  Eq. (\ref{angle-m}) 
can be evaluated as 
$$
\frac{2EV_e}{\Delta m^2_{13}} = 5.1 \cdot 10^{-2} 
\frac{\rho Y_e}{g/cc} \cdot  \frac{E}{1 {\rm GeV}} \cdot
\frac{3 \cdot 10^{-3} {\rm eV}^2}{\Delta m^2} < 0.1
$$   
for the matter density  in the Earth $(3 - 10)$ g/cc. That is,  
the additional rotation is much smaller than the vacuum (1 - 3) 
rotation. The angle $\theta_{13}^m$ can be considered as a 
small correction to $\theta_{13}$: 
\be
\tilde \theta_{13} = \theta_{13} + \Delta \theta_{13}  
\approx \theta_{13} \left(1 + \frac{2EV}{\Delta m^2_{31}}\right). 
\label{theta-t}
\ee
Basically,  $\tilde \theta_{13}$ is the mixing angle in matter 
in the two neutrino approach: 
$\tilde \theta_{13} \approx \theta_{13}^m$,  and  
plus sign in Eq.~(\ref{theta-t}) reflects the fact
that matter enhances the mixing in the neutrino channel for the normal mass hierarchy. 
In the antineutrino channel the sign is negative.  
For the inverted mass hierarchy the sign of $\Delta m^2_{31}$
should change, and  the correction to $s_{13}$  (\ref{theta-t}) is  
positive in the antineutrino channel and negative in the neutrino channel. 

As a consequence of the additional rotation we find the following.   

1. The correction to $H_{11}'$ element appears which can be considered as
the correction to the potential: 
\be
\Delta H_{11} = \Delta V \approx - V s_{13}^2 
\left(1 - \frac{2EV}{\Delta m^2_{13}}\right). 
\label{corr11}
\ee
It can be safely neglected. 

2. The $H_{23}$ and  $H_{32}$ elements  are generated: 
\be
H_{23} = H_{32} = s_{12} c_{12} s_{13} V 
\left(\frac{\Delta m^2_{12}}{\Delta m^2_{13}}\right) 
\label{corr-23}
\ee
which are again negligible.  Also corrections to the $H_{33}'$ 
element can be neglected.

Combining the  rotations introduced above we 
find the projection matrix  $\tilde{U}$ (\ref{p-basis})  
which defines the propagation basis:   
\be
\tilde{U} = U_{23} D_{\delta_{CP}} \tilde{U}_{13} = U_{23} D_{\delta_{CP}} U_{13} U_{13}' 
= 
\left(\begin{array}{ccc}
\tilde{c}_{13}   &  0  & \tilde{s}_{13} \\
- \tilde{s}_{13} s_{23} e^{i\delta_{CP}} &    c_{23}  &
\tilde{c}_{13} s_{23} e^{i\delta_{CP}}  \\
- \tilde{s}_{13} c_{23} e^{i\delta_{CP}} &  - s_{23}   & \tilde{c}_{13} c_{23} e^{i\delta_{CP}}
\end{array}
\right) ~ .
\label{project}
\ee

In the propagation basis the evolution equation 
can be obtained from  Eqs. (\ref{evolution}, \ref{project}) 
\be
i \frac{d \tilde{\nu}}{dt} =
\left(\tilde H + i \frac{d \tilde{\theta}_{13}}{dt}
\hat{\lambda}\right) \tilde{\nu},
\label{evol-prop}
\ee
where 
\be
\tilde{H} \approx
\left(\begin{array}{cc}
H_2   & 0 \\
0  & \Delta m_{31}^2/2E  + V s_{13}^2
\end{array}\right)\, ,
\label{matr2}
\ee
\be
H_2  = \frac{1}{2E} U_{12} M_2  U_{12}^{\dagger} + V c_{13}^2, ~~~~~
M_2 \equiv  diag(0, \Delta m_{21}^2).  
\label{h-2}
\ee 
In (\ref{evol-prop}) the matrix $\hat{\lambda}$ has all zero elements 
but $\lambda_{13} = - \lambda_{31} = 1$. 
The last term in Eq. (\ref{evol-prop}) can be evaluated as 
$d\tilde{\theta}_{13}/dt \sim \Delta \theta_{13} / R_{earth}$ for
trajectories crossing the mantle only. For the core crossing trajectories 
the derivative can be large at the border between the 
core and the mantle. However,  because of averaging of oscillations  
driven by $\Delta m_{31}^2$ we can neglect this dependence too.

According to Eq.~(\ref{matr2}),  the state 
$\tilde{\nu}_{\tau}$  decouples from the rest of the system
and evolves independently. 
The $(\tilde{\nu}_e, \tilde{\nu}_{\mu})$ sub-system 
evolves according to the  2$\times$2 Hamiltonian $H_2$
($\tilde{\nu}_e - \tilde{\nu}_{\mu}$ sub-matrix in Eq.~(\ref{matr2})).  
This Hamiltonian is determined by the solar oscillation parameters 
$\Delta m_{21}^2$, $\tan^2 \theta_{12}$ and the potential 
$V \approx V c_{13}^2$. 
Thus, in the propagation basis the three neutrino problem is reduced 
to two neutrino problem. 

Correspondingly, the evolution matrix S 
(the matrix of   amplitudes) in the propagation basis 
$(\tilde{\nu}_e, \tilde{\nu}_{\mu}, \tilde{\nu}_{\tau})$   
has the following form:
\be  
\tilde{S} \approx 
\left(\begin{array}{ccc} 
\tilde{A}_{ee}   & \tilde{A}_{e \mu}    & 0 \\
\tilde{A}_{\mu e}   & \tilde{A}_{\mu \mu}    &   0    \\
0        & 0         & \tilde{A}_{\tau \tau} 
\end{array}
\right) ~~ ,~~
\label{matr-s}
\ee
where 
\be 
\tilde{A}_{\tau \tau} = \exp(-i\phi_{3})\,, \quad \quad      
\phi_{3} \approx \frac{\Delta m_{31}^2 L}{2E},  
\label{phase}
\ee
and  $L$ is the total distance traveled by neutrinos. 
Other amplitudes in (\ref{matr-s}), $\tilde{A}_{ee}$, $\tilde{A}_{e \mu}$,... 
should  be found by solving 
the two neutrino evolution equation with the Hamiltonian $H_2$.

Since the oscillations driven by $\Delta m_{31}^2$ are averaged out,
the dependence of the   effects on the type of mass hierarchy 
(normal, inverted) appears only in the value of $\tilde{s}_{13}$ for neutrino and 
antineutrino  channel. In what follows we will present results for
normal mass hierarchy.  For inverted hierarchy the difference of
results is very small and  can be largely absorbed in redefinition of $s_{13}$. 

\subsection{Flavor transitions}

\begin{figure}[ht]
\centering\leavevmode
\epsfxsize=0.7\hsize
\epsfbox{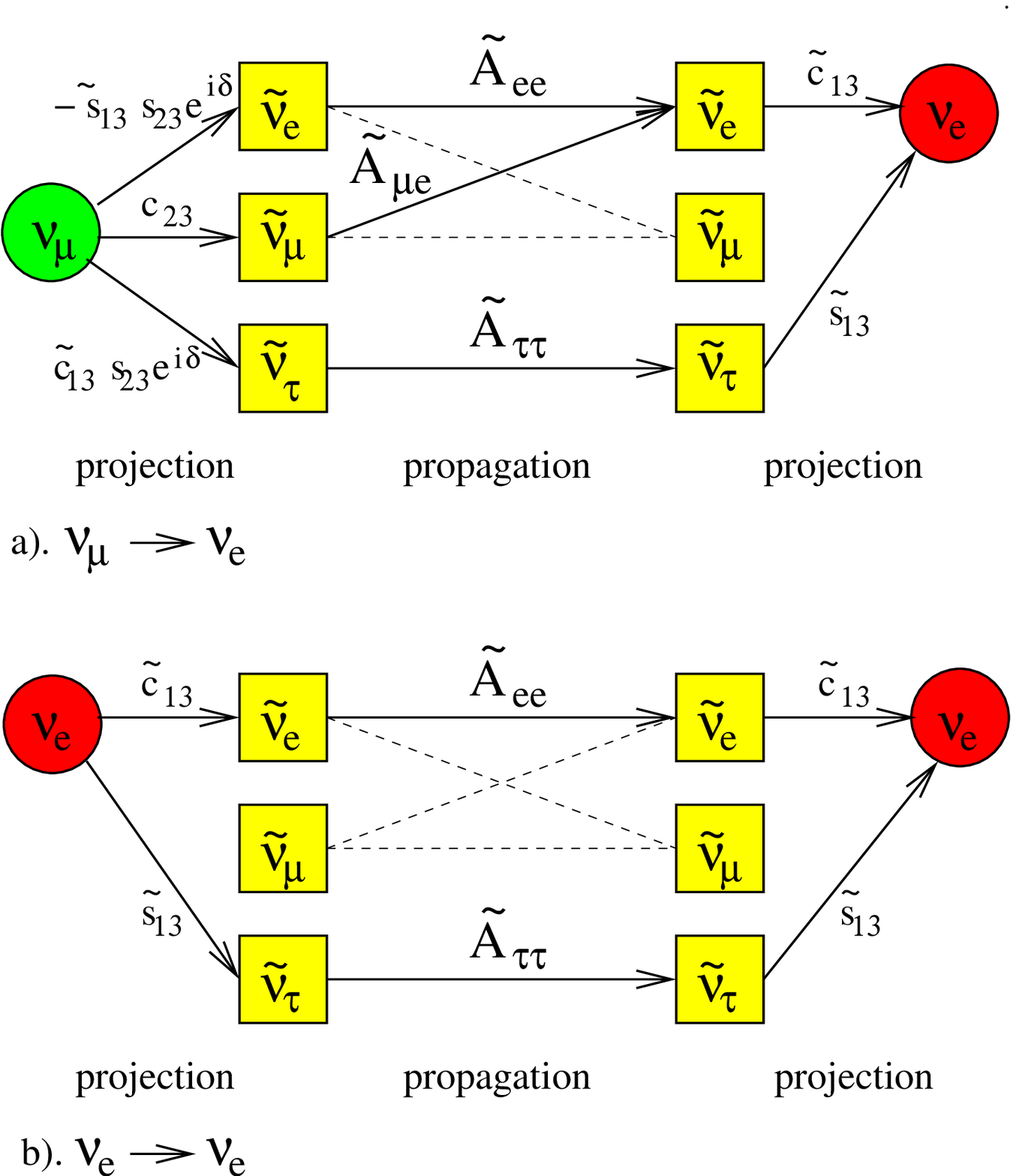}
\caption{The neutrino transition  scheme. Initial and final flavor
states are shown in  circles. In boxes we show the states of the
propagation basis. Lines connect  states between which the transitions
can occur. The lines with arrows indicate  transitions and projections
relevant for oscillation channels of interest.
}
\label{fig1}
\end{figure}

Let us find  the  probabilities of  
the $\nu_{\mu} \leftrightarrow \nu_e$ oscillations, $P_{\mu e}$,  and 
the $\nu_e \leftrightarrow \nu_e$ oscillations,  
$P_{ee}$,  relevant for our problem. 
The calculation proceeds in the three steps 
(see the transition scheme presented in fig.~\ref{fig1}):    
(1) projection of the initial flavor state on to the 
propagation basis, (2) evolution in the propagation basis, 
(3) projection of the result of the evolution in the 
propagation basis on to the final flavor state.  
According to this picture the $S-$matrix  in the flavor basis 
equals: 
\be
S = \tilde{U} \tilde{S} \tilde{U}^{\dagger},     
\label{s-fl}
\ee          
where $\tilde{U}$ and  $\tilde{S}$ are given by Eqs. (\ref{project}) 
and (\ref{matr-s}). 

Using Eqs.  (\ref{s-fl},\ref{matr-s},\ref{project}) we find 
\be
P_{\mu e} = \left| 
-\tilde{s}_{13} \tilde{c}_{13} s_{23} \tilde{A}_{ee} e^{i\delta_{CP}} + 
\tilde{c}_{13} c_{23} \tilde{A}_{\mu e} 
\right|^2 +
\tilde{s}_{13}^2 \tilde{c}_{13}^2 s_{23}^2.  
\label{mue-pr}
\ee
For the sub-GeV sample the oscillations driven by $\Delta m_{31}^2$ 
are averaged out, so that there is no interference effect due to 
state $\tilde{\nu}_{\tau}$. At the same time, 
according to (\ref{mue-pr}) 
the amplitudes $\tilde{A}_{ee}$ and $\tilde{A}_{\mu e}$ 
interfere. It is this interference  which produces effect
we interested in this paper. Notice that amplitudes
$\tilde{A}_{ee}$ and $\tilde{A}_{\mu e}$ are   
both due to the solar oscillation parameters.  However, their interference appears 
due to presence  of the third neutrino (non-zero $s_{13}$). 
In the limit $s_{13} = 0$ the interference disappears. 
In what follows we will call the interference of the amplitudes 
(with solar oscillation parameters) due to non-zero $U_{e3} \sim s_{13}$ 
as the {\it $U_{e3}$ induced interference}.  
 
According to (\ref{mue-pr}), there is no interference 
of the amplitudes driven by the atmospheric, $\Delta m_{31}^2$, 
and solar $\Delta m_{21}^2$ mass splittings. This interference is averaged
out for the most part of the zenith angles. 
If  $\cos\Theta_{\nu}  > 0$  (above the horizon),   neutrinos propagate
in the atmosphere, where the matter effect can be neglected.  
The effect of corresponding  interference terms is 
very small:  below (0.2 - 0.3) \% (see Appendix), though  we take it  into 
account in our numerical calculations.

The probability  (\ref{mue-pr}) can be written explicitly as 
\be
P_{\mu e} = 
\tilde{c}_{13}^2 c_{23}^2 P_2  
- 2 \tilde{s}_{13} \tilde{c}_{13}^2 s_{23} c_{23} (\cos \delta_{CP} R_2 -
\sin \delta_{CP} I_2) 
+ \tilde{s}_{13}^2 \tilde{c}_{13}^2 s_{23}^2 (2 - P_{2}), 
\label{mue-pr2}
\ee  
where 
\be
P_2 \equiv |\tilde{A}_{\mu e}|^2  =  1 - |\tilde{A}_{ee}|^2,~~~ 
R_2 \equiv Re(\tilde{A}_{\mu e}^*\tilde{A}_{ee}), ~~~~
I_2 \equiv Im(\tilde{A}_{\mu e}^*\tilde{A}_{ee})
\ee
are the $2\nu$  probabilities in the propagation basis. 

Similarly, we  get  $P_{ee}$: 
\be
P_{ee} = \tilde{c}_{13}^4 (1 - P_2) + \tilde{s}_{13}^4.   
\label{ee-pr}
\ee
No induced interference appears here due to zero projection of  
$\nu_e$  on to $\tilde{\nu}_{\mu}$ state (see (\ref{project})). 

For antineutrinos, the probabilities $\bar{P}_2$, $\bar{R}_2$, $\bar{I}_2$ 
should be obtained  by  replacement
of $V \rightarrow -V $ in the Hamiltonian $H_2$  of Eq.~(\ref{matr2}), 
and the sign of phase  $\delta_{CP}$ should be changed:  
\be 
\overline{P}_2 = P_2(-V_e),~~~ \overline{R}_2 = R_2(-V_e),~~~
\overline{I}_2 = I_2(-V_e),~~~
~~~~ \overline{\tilde{s}}_{13} = \tilde{s}_{13}(-V_e), ~~~ \delta_{CP}
\rightarrow - \delta_{CP}.  
\ee
As a result, 
\be
\bar {P}_{\mu e} =
\overline{\tilde{c}}_{13}^2 c_{23}^2 \bar{P}_2
- 2 \overline{\tilde{s}}_{13} \overline{\tilde{c}}_{13}^2 s_{23}
c_{23} (\cos \delta_{CP} \bar{R}_2 +  \sin \delta_{CP} \bar{I}_2
+ \overline{\tilde{s}}_{13}^2 \overline{\tilde{c}}_{13}^2 s_{23}^2 (2
- \bar{P}_{2}), 
\label{barmue-pr2}
\ee
\be
\bar{P}_{ee} = \overline{\tilde{c}}_{13}^4 (1 -\bar{P}_2) +
\overline{\tilde{s}}_{13}^4.
\label{ee-prbar}
\ee

\begin{figure}[ht]
\centering\leavevmode
\epsfxsize=0.7\hsize
\epsfbox{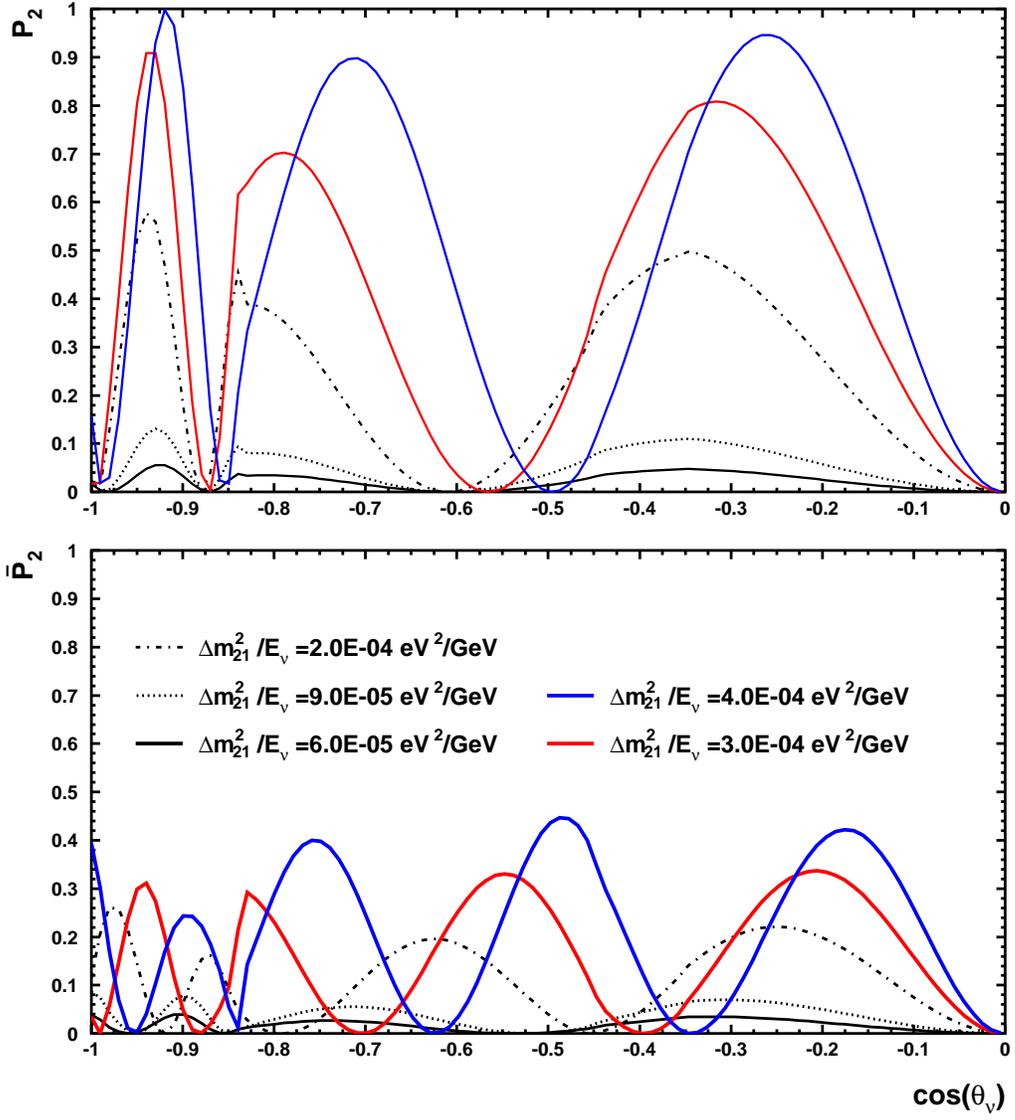}
\caption{
Zenith angle dependences of the transition probabilities    
for neutrinos,  $P_2$,  
(upper panel), and for antineutrinos (lower panel), $\bar{P}_2$,  
for different values of  $\Delta m^2_{12}/E$ and  $\sin^2 2
\theta_{12} = 0.82$. 
}
\label{fig2}
\end{figure}

\begin{figure}[ht]
\centering\leavevmode
\epsfxsize=0.7\hsize
\epsfbox{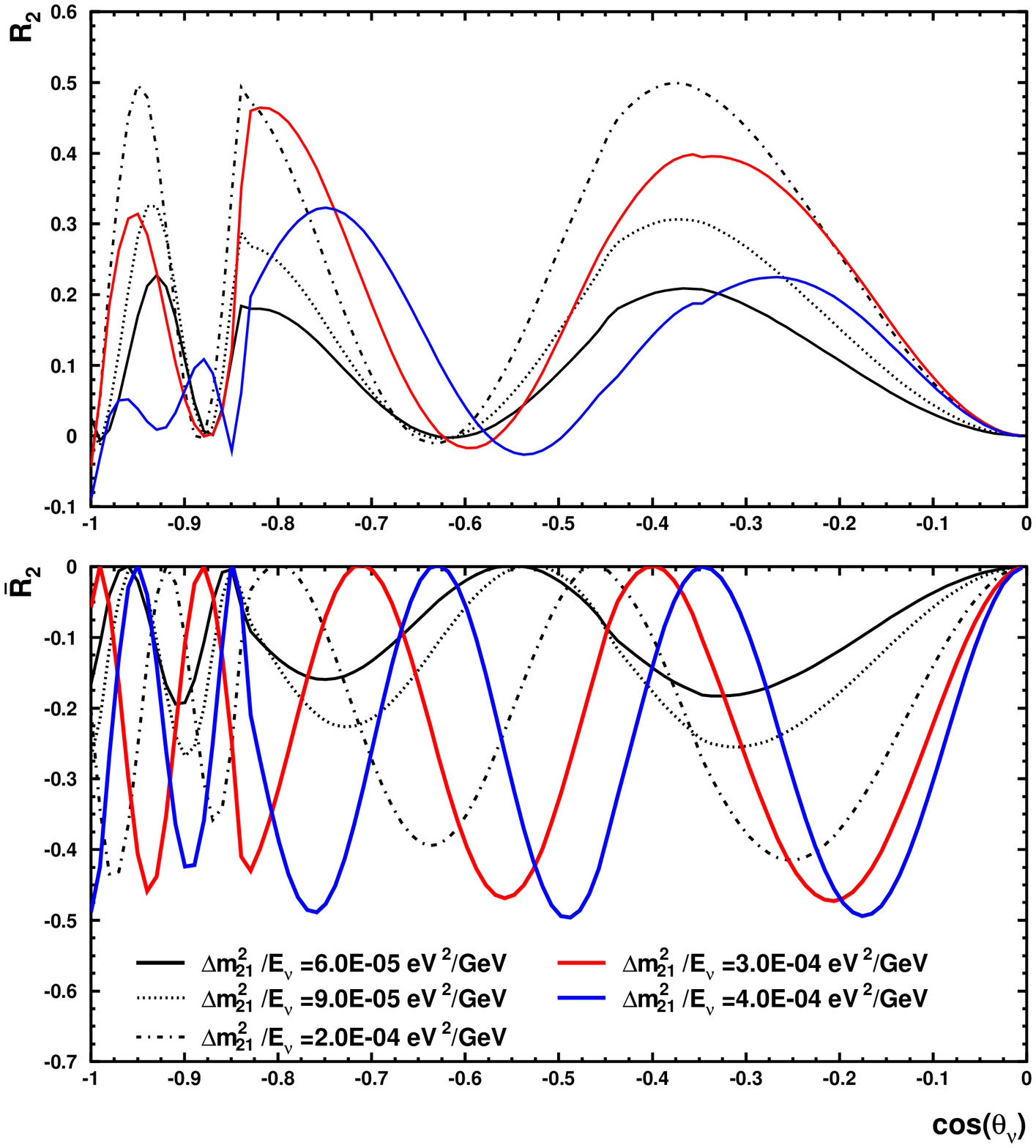}
\caption{
Zenith angle dependences of the real part of the  interference probability   
for neutrinos, $R_2$ (upper panel),  and 
for antineutrinos, $\bar{R}_2$  (lower panel),  for different values 
of $\Delta m^2_{12}/E$ and $\sin^2 2 \theta_{12} = 0.82$.
}
\label{fig3}
\end{figure}

Let us consider the  two neutrino  probabilities
${P}_2$,  ${R}_2$, ${I}_2$ as well as $\bar{P}_2$, $\bar{R}_2$, $\bar{I}_2$ in details. 
We have calculated them numerically (see results in the
figs.~\ref{fig2} - \ref{fig4}) 
using the distribution of density in the Earth from Ref.~\cite{earthmodel}.

Properties of the probabilities can be well
understood using their expressions  in medium  with constant density:  
\be
P_2 = \sin^2 2\theta_{12}^m  \sin^2 \frac{\phi_m}{2}~,
\label{const-P}
\ee
\be
R_2  = - \sin 2\theta_{12}^m \cos 2\theta_{12}^m \sin^2 \frac{\phi_m}{2}~,
\label{const-R}
\ee
\be
I_2 =  - \frac{1}{2}\sin 2\theta_{12}^m \sin \phi_m. 
\label{const-I}
\ee
Here  $\phi_m = \Delta H_{12} R_{earth} \cos \Theta_{\nu}$, 
is the phase of oscillations in matter, where $\Delta H_{12}$ 
is the difference of the eigenvalues of the Hamiltonian in matter,
$R_{earth}$ is the Earth  radius and $\Theta_{\nu}$ 
is the zenith angle of neutrinos. In (\ref{const-P} - \ref{const-I})
$\theta_{12}^m$ is  the 1-2 mixing in matter  determined by  
\be 
\sin 2\theta_{12}^m = 
\frac{\sin 2\theta_{12}}{\cos^2 2\theta_{12}(1- E_{\nu}/E_R)^2 +  \sin^2 2\theta_{12}}. 
\ee
The resonance neutrino energy equals 
\be
E_R \approx \frac{\Delta m^2_{21} \cos 2\theta_{12}}{2V  c^2_{13}} =
0.238 ~{\rm GeV} 
\left( \frac{\Delta m^2_{21}}{7 \cdot  10^{-5}{\rm eV}^2 } \right)
\left( \frac{2.0 {\rm g/cm^3}}{Y_e\rho} \right)
\cos 2\theta_{12} .
\ee
In the mantle, for the present best fit  value 
$\Delta m^2_{21} = 7.3 \cdot 10^{-5}$~eV$^2$ 
and for $\sin^2 2\theta_{12}=0.8$ we get  $E_R~=~0.10$~GeV
which is below the threshold of  sub-GeV range. 
Therefore  for  $\Delta m^2_{21} \sim  (5 - 7) \cdot 10^{-5}$ eV$^2$ 
and $E_{\nu} \sim (0.1 - 0.5)$ GeV  the oscillations occur  
in the matter dominated regime when the potential is 
larger than the kinetic term: $V >  \Delta m^2 / 2E$.

For $\Delta m^2_{21}/E_{\nu}  < 10^{- 4}$~eV$^2$/MeV 
the depth of oscillations is roughly proportional to 
$(\Delta m^2)^2$.  The oscillation length, $l_m$, is close to the
refraction length, $l_0$, and only weakly depends on energy: 
\be
\sin^2 2\theta_m  \sim  \sin^2 2\theta_{12} 
\left(\frac{\Delta m^2_{21}} {2E_{\nu} V} \right)^2~,~~~~~~  
l_m \approx l_0 = \frac{2\pi}{V}. 
\label{mdominate}
\ee

With increase of  $\Delta m^2/E$, the mixing parameter  
$\sin^2 2\theta_m$, and consequently,   
$P_2$ approach  1  in the resonance in the neutrino channel. 
In the antineutrino channel the mixing and $\bar{P}_2$ increase  but they are  always
below vacuum values. 

The propagation (at least in the mantle of the Earth) has a character of
oscillations with quasi constant depth and length. 
Correspondingly, $P_2$,   $R_2$ and $I_2$ have an oscillatory 
behavior with $\cos \Theta_{\nu}$. 

In Fig.~\ref{fig2} (upper panel) we show  dependence of $P_2$  
on the zenith angle of neutrino, $\Theta_{\nu}$,   
for different values of $\Delta m^2_{21}/E$. 
The depth of oscillation of $P_2$  is determined basically by $\sin^2
2\theta_{12}^m$. ${P}_2$ monotonously increases with $\Delta m^2_{21}$. 
Notice that the first oscillation  maximum is achieved 
at $\cos \Theta_{\nu}~\sim -0.35 \div  - 0.4$ and the effect is zero at $\cos
\Theta_{\nu} \sim - 0.64$.  Second maximum is for the trajectories 
at the border between core and mantle: 
$\cos \Theta_{\nu} =   - 0.84$. 
For $\cos \Theta_{\nu} <  - 0.84$ neutrinos
cross both the  mantle and  core of the Earth. The  interplay of
the oscillations in the mantle and in the core leads to some
enhancement of the transition  probability  in spite of larger density
of the core.  For core crossing trajectories the 
period of oscillation is smaller.


For antineutrinos (fig.~\ref{fig2}, bottom panel) the mixing angle is suppressed.  
The oscillation length is smaller than $l_0$.  
With increase of $\Delta m^2/E$  the mixing (depth of 
oscillations) increases whereas  the oscillation length decreases approaching
vacuum  values.  

The oscillation effects in the antineutrino channel 
are  smaller by factor 2 - 3. 

$R_2$ has similar oscillatory  dependence on $\cos \Theta_{\nu}$
(fig.~\ref{fig3}) with the depth of oscillations  given by  $\sim \sin
2\theta_{12}^m \cos 2\theta_{12}^m$   (see Eq. (\ref{const-R})).   In
contrast to $P_2$,  with increase of $\Delta m^2_{21}$ 
the real part,  $R_2$,  first increases, reaches maximum  at 
$\Delta m^2_{21} \sim 7 \cdot 10^{-5}$ eV$^2$   
($\Delta m^2_{21}/E_{\nu}  \sim 2 \cdot 10^{-4}$   eV$^2$/GeV )
and then decreases. The interference term is zero in the resonance:  
$\Delta m^2_{21}/E_{\nu}  \sim 7 \cdot 10^{-4}$   eV$^2$/GeV.  
It changes the sign with further increase of $\Delta m^2_{21}/E_{\nu}$
approaching vacuum value.

In general (without rely on constant density approximation) the real part of the 
interference  term  can be written as 
\be
R_2 = Re(\tilde{A}_{ee}\tilde{A}_{\mu e}^*) = 
\sqrt{P_2 (1 - P_2)} 
\cos(\phi_{ee} - \phi_{\mu e}),
 \label{reeexp}
\ee
where $\phi_{ee} \equiv arg(\tilde{A}_{ee})$ 
and $\phi_{\mu e} \equiv  arg(\tilde{A}_{\mu e})$. 
From  Eq. (\ref{reeexp}) we conclude  that maximal value  equals 
$R_2^{max} = \frac{1}{2}$.
It corresponds to  $P_2 = 1/2$ and 
$\phi_{ee} =  \phi_{\mu e} + \pi k$, ($k = integer$).  
For the sub-GeV sample we find that $P_2 = 1/2$ 
is achieved at $\Delta m_{21}^2 \sim  7 \cdot 10^{-5}$ eV$^2$, that is,  
for the present best fit value.

In the constant density approximation the phase factor equals :
\be
\cos(\phi_{ee} - \phi_{\mu e}) = 
- \frac{\sin \phi_m \cos 2\theta_{12}}
{\sqrt{1 - \sin^2 \phi_m \sin^2 2\theta_{12}}}~. 
\label{cos}
\ee   
From this equation we find  that in  maximum  of the oscillation probability 
($\sin \phi_m = 1$):   $\cos(\phi_{ee} - \phi_{\mu e}) = 1$,  
and consequently,  the interference term reaches maximum.

According to (\ref{const-P}, \ref{const-R}) 
\be
\frac{P_2}{R_2} \approx \tan \theta_{12}^m, 
\ee
and therefore the interference probability 
dominates at high energies or low $\Delta
m^2_{21}$, when $2\theta_m < \pi/4$. The latter  corresponds to
$\Delta m^2_{21}/E_{\nu}  \sim 2 \cdot 10^{-4}$   eV$^2$/GeV 
for the mantle of the Earth.  
Thus,  for  $\Delta m^2_{21} \sim 7 \cdot 10^{-5}$ eV$^2$,  
$P_2$ and $R_2$  are comparable.  For larger $\Delta m^2_{21}$, the LMA probability  $P_2$ 
dominates, whereas for smaller $\Delta m^2_{21}$,   
the interference probability $R_2$ is larger. 

The interference term $R_2$  has opposite sign  for neutrinos and
antineutrinos (fig.~\ref{fig3},  bottom panel) due to change of the sign of $V$.  
This result can be easily understood using the  
constant density approximation (\ref{const-R}).  
Indeed, the mixing angle in matter, $\theta_{12}^m$,   
differs for neutrino and antineutrino. For definiteness,  let us assume that 
vacuum mixing angle is below $\pi/4$, as is favored by the present 
solar neutrino data.   In this case matter suppresses the mixing in the 
antineutrino channel,  and enhances mixing in the neutrino channel.  
So, we have  $\theta_{12}^m (\bar{\nu}) < \theta_{12} < \pi/4$, and 
$\theta_{12}^m (\nu) > \theta_{12}$. Furthermore, 
for $\Delta m^2 < 10^{-4}$ eV$^{2}$ (where the interference effect
is large) and for neutrino energies  relevant for the sub-GeV sample, 
the mixing is above resonance:  $\theta_{12}^m (\nu) > \pi/4$. 
Therefore $\cos 2\theta_{12}^m$ is positive for antineutrinos and negative 
for neutrinos, and since  $\sin 2\theta_{12}^m$ is positive in both
channels the interference term has opposite 
sign for neutrinos  and antineutrinos.  

Also behavior of the interference term $\bar{R}_2$ in the antineutrino 
channel  with energy  differs from that of $R_2$. 
In the antineutrino channel $2\theta_{12}^m (\nu)$ increases. 
Correspondingly, $\bar{R}_2$ reaches maximum when $2\theta_{12}^m (\nu) = \pi/4$ 
($\Delta m^2_{21}/E_{\nu}  \sim 4 \cdot 10^{-4}$   eV$^2$/MeV)   and
then it decreases.


The imaginary part, $I_2$,  (fig.~\ref{fig4})  changes the sign with increase of the 
oscillation  phase,  and consequently, with $\cos \Theta_z$. 
So, integration over the zenith angle leads to strong suppression of 
$I_2$, and therefore, the  CP-violating effects. The depth of oscillations 
increases  according to $\sin 2\theta_{12}^m/2$ and   maximal value,
$I_2 = 1/2$,  is achieved  in the resonance, $\Delta m^2_{21}/E_{\nu}  \sim 7 \cdot
10^{-4}$   eV$^2$/GeV.

\subsection{Neutrino fluxes in presence of oscillations}

Let  $F_e^0$ and $F_{\mu}^0$ be the electron and muon neutrino fluxes at 
the detector in the absence of  oscillations. Then, 
the flux with oscillations  can be written as 
\be
F_e = F_e^0 P_{ee} +  F_{\mu}^0 P_{\mu e} = 
F_e^0 ( P_{ee} + r  P_{\mu e}),    
\label{fluxe-osc}
\ee
where 
$$
r(E, \Theta_\nu) \equiv \frac{F_{\mu}^0(E, \Theta_\nu)}{ F_e^0(E,
\Theta_\nu)} 
$$ 
is the ratio of the original fluxes.  In the sub-GeV range the  ratio
$r$ depends both on the zenith angle and on the neutrino 
energy rather weakly and can be approximated  by $r = 2.04 - 2.06$.

Inserting the probabilities 
$P_{ee}$ and  $P_{\mu e}$ from Eqs. (\ref{ee-pr}) and 
(\ref{mue-pr2})  in  Eq. (\ref{fluxe}) we get expression 
for the  relative change  of the $\nu_e-$flux:
$$
\frac{F_e}{F_e^0} -  1 =  (r c_{23}^2 - 1) P_2 - ~~~~~~~~~~~~~~~~~~~~~~~~~~~~~~~~~~~
$$
$$- r \tilde{s}_{13} \tilde{c}_{13}^2 \sin 2\theta_{23} 
(\cos\delta_{CP}~ R_2  - \sin \delta_{CP} ~I_2) - ~~~~~~~~~~~~~~~~~~~~
$$
\be
-2 \tilde{s}_{13}^2 (1 - r s_{23}^2)  
- \tilde{s}_{13}^2 P_2 (r - 2) 
+ \tilde{s}_{13}^4 (1 - r s_{23}^2) (2 - P_2).  
\label{fl-excess}
\ee

Let us consider the   terms of  this equation in order. 

The first term on the right hand side (zero order in
$s_{13}$) corresponds to the  LMA contribution we have discussed in \cite{orl1}. 
Being proportional to $P_2$ 
this term  increases with  $\Delta m_{21}^2/E$  
up to  the resonance value 
$\Delta m_{21}^2/E = 7 \cdot 10^{-4}$ eV$^2$/GeV, where $P_2  \sim 1$.  
The probability is screened by  the  factor $(r c_{23}^2 - 1)$.  
Since  $r \approx  2$ it leads to excess of the flux for $\theta_{23} < 45^\circ$ 
and to deficiency for $\theta_{23} >  45^\circ$.
For  $\theta_{23} = 45^\circ$ the screening factor equals 0.02 - 0.03. 
This term does not depend on $s_{13}$. \\

The second term in (\ref{fl-excess}) is the effect of  induced interference. 
It has the following properties.  

\begin{itemize}

\item
The term  depends on $s_{13}$ linearly and therefore  
its effect may not be strongly suppressed even for small 
$s_{13}$. The interference depends  on the  sign of $s_{13}$. 

\item
The interference term does not have  screening factor, so it can dominate 
for 2-3 mixing close to maximum. 
Its smallness is mainly due to smallness of $s_{13}$ as well as  $R_2$ and $I_2$.

\item
The interference term is proportional to $\sin 2\theta_{23}$ and therefore it is 
sensitive to the sign of $\theta_{23}$.  

\item
With increase of $\Delta m_{21}^2/E$ the real part (similarly to
$R_2$) first increases,  reaches maximum at 
$\Delta m_{21}^2/E = 2 \cdot 10^{-4}$ eV$^2$/GeV, and then decreases
and changes the sign in the  resonance. The imaginary part  increases
up to the resonance value of $\Delta m_{21}^2/E$ where  $I_2^{max}  = 1/2$. 

\item
For antineutrinos the interference term (as $\bar{R}_2$) 
has the opposite sign with respect to the neutrino term. 
The amplitudes of the imaginary part 
have the same sign for neutrinos and antineutrinos.  
$\bar{R}_2$ reaches value 1/2 at higher 
$\Delta m^2_{12}/E$ than ${R}_2$ does. \\

\end{itemize}

Last three terms in Eq. (\ref{fl-excess}) are of the order
$\tilde{s}_{13}^2$ or of higher power of $\tilde{s}_{13}$. Practically
among these terms only the first one can give significant 
contribution provided that the (2 - 3) mixing deviates from  maximum. 
This term does not  depend on 
$\Delta m^2_{12}/E$. 
Besides $s_{13}^2$ suppression the second term has 
an additional small factor $(r - 2)$. Its contribution does not exceed 0.1\%.  
The third term is proportional to  $\tilde{s}_{13}^4$ and also
contains screening factor. 

For exactly maximal 2-3 mixing and  $r = 2$ we get from
(\ref{fl-excess}): 
\be
\frac{F_e}{F_e^0} -  1 =  - r \tilde{s}_{13} \tilde{c}_{13}^2 
(\cos \delta~ R_2 - \sin \delta~ I_2). 
\label{fl-excessM}
\ee  
That is, only the interference term gives a contribution. 
Since  in the sub-GeV sample $r = 2.04 - 2.06$, 
no complete cancellation is possible. 

In what follows we will  describe the deviation of the 2-3 mixing by the parameter
\be
D_{23} \equiv \frac{1}{2} -s^2_{23}. 
\label{dev}
\ee
From the $2\nu$ analysis of the atmospheric neutrino data (\ref{atmdata}) we get 
\be
|D_{23}| < 0.15,~~~~~ (90 \% C.L.).
\label{devb}
\ee
Note that for consistency such a  bound should be obtained from the $3\nu$ analysis 
which includes the LMA oscillations.

\section{Oscillation effects in  the  $e$-like  events}

In what follows we  calculate  dependences of the number  of
$e-$like events on the zenith angle of electron, $\Theta_e$, and the
electron energy. The general expression for the number of e-like
events, $N_{e}$, as a function of $\Theta_e$ is
\begin{eqnarray}
N_{e} \propto \sum_{\nu\overline{\nu}}
 \int  & \left. dE_\nu dE_e d(\cos\Theta_\nu) dh  \; 
F_e(E_{\nu},\Theta_{\nu}) \displaystyle{\frac{d{ \sigma}}{dE_e}}
 \;\times \right.  \nonumber \\ 
 & \times \; \Psi(\Theta_e,\Theta_{\nu},E_{\nu})   
\kappa_e(h,\cos\Theta_\nu,E_\nu)   
\varepsilon(E_e)  \;\;,
\label{event1}
\end{eqnarray}
where $F_e$ is the atmospheric $\nu_e$-flux
at the detector given in Eq.~(\ref{fluxe-osc}) (the fluxes 
$F_e^0$ and $F_{\mu}^0$ without oscillations are taken from 
Ref.~\cite{flux});
$d\sigma / dE_{e}$ are the differential cross
sections taken from  Ref.~\cite{LLS},  $\kappa_e$ is the normalized
distribution of neutrino production points, $h$ is the height of
production, $\varepsilon(E_e)$ is the detection efficiency of the
electron, $\Psi$ is the ``dispersion'' function which describes
deviation of the lepton zenith angle from the neutrino 
zenith angle~(~for details see Ref.~\cite{compute}).

The integration over the neutrino zenith angle and
neutrino energy leads to  significant smearing of the 
$\Theta_e$ dependence.   
The average angle between the neutrino and the outgoing charged lepton  is
about $60^{\circ}$ in the sub-GeV range. 
Furthermore, neutrinos and antineutrinos of a given flavor are not distinguished 
in the present atmospheric neutrino experiments, 
so that  their  signals are summed in 
Eq.~(\ref{event1}) which  leads typically to further weakening of the
oscillation effect.

According to Eqs. (\ref{fl-excess})  and (\ref{event1}) the relative
change  of the $e-$like events, can be represented as  the sum of three contributions: 
\be 
\epsilon_e \equiv \frac {N_e}{N_e^0} - 1 =
\epsilon_e^{LMA} + \epsilon_e^{int} + \epsilon_e^{Ue3},  
\label{cont}
\ee
where 
\be 
\epsilon_e^{LMA}  \approx  (r  c_{23}^2 - 1)  
[(1 - \xi) \langle P_2 \rangle + \xi \langle \bar{P}_2 \rangle ]
= (r  D_{23}    + 0.5 r - 1)  
[(1 - \xi) \langle P_2 \rangle + \xi \langle \bar{P}_2 \rangle ]
\label{eLMA}
\ee  
is  the contribution of oscillations driven by  the solar (LMA) parameters, 
\be
\epsilon_e^{int} = \cos \delta~ \epsilon_R^{int} - \sin \delta~
\epsilon_I^{int}, 
\ee
\be
\epsilon_R^{int} \approx -  r 
\langle \tilde{s}_{13} \tilde{c}_{13}^2\rangle  
\sin 2 \theta_{23} 
\left[(1 - \xi) \langle R_2 \rangle + \xi \langle \bar{R}_2 \rangle \right]
\label{eps-int}
\ee 
\be
\epsilon_I^{int} \approx - r \langle \tilde{s}_{13}
\tilde{c}_{13}^2\rangle  
\sin 2 \theta_{23} \left[(1 - \xi) \langle I_2 \rangle - \xi \langle
\bar{I}_2 \rangle \right], 
\label{eps-int1}
\ee 
is the interference term,    
\be
\epsilon_e^{Ue3} =  - 2 \langle \tilde{s}_{13}^2 \rangle
(1 - r  s_{23}^2)  
\label{eps}
\ee 
is the $U_{e3}$-induced term. Here 
 $\langle X \rangle \equiv \langle  X  (E_e, \Theta_e) \rangle$ 
is the  quantity $X$  
averaged over appropriate energy and zenith angle intervals of neutrino as well as
final lepton (\ref{event1});  they are functions of the electron energy $E_e$ 
and the zenith angle $\Theta_e$.  
Here we have summed up the effect of neutrinos and antineutrinos, 
assuming that the detector does not 
identify the electric charge of the lepton. The parameter 
\be
\xi \equiv \frac{\bar{N}_0}{\bar{N}_0 + N_0} = 0.3
\ee
describes the relative contribution of the antineutrino flux without oscillations. 
In estimations one can take $\langle E \rangle = 0.4$ GeV as an
effective neutrino energy  relevant  for the sub-GeV sample of events.

Properties of different contributions (\ref{cont}) 
reproduce basically the properties of the corresponding 
terms in the expression for the flux change (\ref{fl-excess}). 
New features of $\epsilon_e$ are  related to the 
strong averaging effect  due to integration over the neutrino energy  
and zenith angle as well as  due to  summation of the neutrino and
antineutrino signals.

In what follows  we take  the experimental data, fluxes, features of detection 
from Ref.~\cite{super-data-used}. 
Recently  Super-Kamiokande collaboration 
has published results of the  refined  analysis of the data (see {\it e.g.},~\cite{sk-eps2003}).  
In particular,  new  detector
simulation, data analysis, input  atmospheric neutrino
fluxes and cross sections have been used.   
Unfortunately, published till now information  
is not enough  to  update our calculation. 
At the same time, we expect that impact of these changes 
on our results will not be significant.

\subsection{The LMA contribution}

Let us assume that $s_{13}$ is zero or very small, so that 
\be 
\epsilon_e \approx \epsilon_e^{LMA} 
\approx r  D_{23}   
\langle P_2 \rangle_{\nu \bar\nu}, 
\ee
where $\langle P_2 \rangle_{\nu \bar\nu} \equiv 
[(1 - \xi) \langle P_2 \rangle + \xi \langle \bar{P}_2 \rangle ]$. 

In Fig.~\ref{fig5}   we show the zenith 
angle dependences of the relative excess of the $e$-like  events 
for different values of $\Delta  m^2_{12}$. 
The upper panel corresponds to large deviation of the 2-3 mixing from 
maximum, so that  the screening factor equals 0.33. Increase of 
the excess with $\Delta m^2_{12}$ follows the dependence of $\langle P_2 \rangle$
and proceeds according to increase of 
$\sin^2 2\theta_{12}^m$. The zenith angle appears due to oscillations of high energy part of 
the sample.   For the best fit point of the LMA solution 
the excess is (3 - 4)\%.

The integrated over the zenith angle excess (which corresponds to 
the result of the third zenith angle bin) can be estimated in the following way. 
At $\Delta m^2_{21} \sim 7 \cdot 10^{-5}$ eV$^2$ 
and  $\sin^2 2\theta_{12} \sim 0.82$ 
the effective mixing parameter $\sin^2 2\theta_{12}^m \approx 0.35$. 
This parameter determines the depth of oscillations of 
$P_2$. The averaging over the
zenith angle gives  $\langle  P_2 \rangle \sim 0.11$. 
(Notice that we average here over the whole interval 
$\cos \Theta_{\nu} = -1 \div +1$ and the oscillation effect for the
down-going neutrinos is small.) 
The averaging over the neutrino and antineutrino fluxes 
leads to  $\langle {P}_2 \rangle_{\nu \bar\nu} \sim 0.09$. So, 
$\epsilon_e \approx 0.09 \times 0.33 = 0.03$ in agreement with 
exact calculations.

For large $\Delta m^2_{12}$ the oscillations can explain the 
experimental results without additional renormalization of the 
original neutrino flux. For smaller $\Delta m^2_{12}$, the data points can 
be  reproduced as a sum of the effects of oscillations and flux renormalization. 

For $\sin^2\theta_{23} = 0.45$ ($\sin^2 2\theta_{23} = 0.99$)  the screening 
is much stronger:  0.127. 
Since the dependence of the excess on $\sin^22\theta_{23}$ factors out,  
the excess  scales as the screening factor: 
the increase of 2-3 mixing leads to decrease of the excess by the 
2.6 (see fig.~\ref{fig5} bottom panel).  This effect can be seen also 
in fig.~\ref{fig6}  where we show the zenith angle dependence of the ratio of events 
with and without oscillations.   

For $\sin^2\theta_{23} > 0.5$ the oscillations produce a deficiency of the 
$e$-like events. The histograms are nearly mirror reflection with respect to 
$N^{osc}_e/N^0_e \approx 1$. 
According to fig.~\ref{fig6} the present data disfavor values $\sin^2
\theta_{23} > 0.5$  which lead to deficit of the $e$-like
events. 
Thus, for $\Delta m^2_{12} = 7.3  \cdot 10^{-5}$ eV$^2$ we find the bound 
$D_{23}  < 0.1$ (without renormalization of the original fluxes).  
It corresponds to a situation when all experimental points 
with error bars in fig.~\ref{fig6}  are above the predicted curve. 
Let us remind that the recent  cosmic ray data 
tend to decrease  the original  neutrino fluxes which strengthens the bound.   
For $\sin^2 \theta_{23} < 0.5$ the bound on deviation from maximal mixing is 
substantially weaker: $D_{23} < 0.4$ (or $\sin^2 2\theta_{23} >  0.36$). 
It   corresponds to  a situation when all experimental points 
with error bars in fig.~\ref{fig6} are below the predicted curve.

The dependence of the excess integrated over the zenith angle   
on the electron energy is shown in fig.~\ref{fig7}.
The excess increases with decrease of energy according to  
increase of $\sin 2\theta_{12}^m$ or $P_2$ (fig.~\ref{fig2}).  
In the very low energy bin, $E < 0.25$ GeV,  
the excess can reach  5 - 6\% for the best fit point. 
The LMA contribution and $\sim 3\%$ 
renormalization of the  flux can give good  description of the data. 
The excess increases from high energies to low energies by about 6\%. 
Therefore  measurements of the  energy dependence with accuracy $\sim 2 \%$ will 
allow to establish existence of the LMA contribution.

With change of the vacuum 1-2 mixing the probability $P_2$ changes only very weakly. 

The oscillation effect depends mainly on $\Delta m^2_{12}$ and  $\sin^2\theta_{23}$. 
After precise determination of the $\Delta m^2_{12}$ (KamLAND will reach  10\% 
accuracy and also SNO will contribute) one can use data on 
$N^{osc}_e/N^0_e$ in atmospheric neutrinos to search for deviation of 2-3 mixing 
from  maximal. In fig.~\ref{fig8} we show contours of constant relative change of 
the $e$-like events in $\cos^2\theta_{23} - \Delta m^2_{12}$ plane. 
Notice that the lines are not symmetric with respect to $\cos^2\theta_{23} = 0.5$ 
due to deviation of $r$ from 2. For this reason the bound from the side 
$\cos^2\theta_{23} < 0.5$ is stronger. 
We assume here that $s_{13} \approx 0$. 
Uncertainties  due to unknown values of $s_{13}$ and $\delta_{CP}$ will be discussed 
in sect IVC.

\subsection{$U_{e3}$ induced interference}

Let us assume that 1-3 mixing is non-zero but $\delta_{CP} = 0$ (or $\pi$). 
Now all three terms in (\ref{cont}) give  contributions to the oscillation 
effect.  
The interference term contribution is determined by the real part $R_2$. 
It  dominates if the  2-3 mixing is  close to maximal. 
In our estimations below we will use 
$\Delta m^2_{12} = 7.3 \cdot 10^{-5}$ eV$^2$ and $\sin^2 2\theta_{12} \sim 0.82$.

In fig.~\ref{fig9} (upper panel) we show the zenith angle distribution of the 
total oscillation effect for different values of $s_{13}$ and $s^2_{23} = 0.45$. 
The LMA contribution is positive and relatively small: its value averaged over the 
zenith  angle equals
$\epsilon_e^{LMA} = 0.8 \%$   
(it is given by the line $s_{13} = 0$, see the upper panel). 
The $U_{e3}$ contribution is also suppressed by the screening factor.  
It is negative for  $s^2_{23} < 0.5$, and 
being quadratic in $s_{13}$, does not depend on the sign of 1-3 mixing. 
We  obtain  
\be
\epsilon^{Ue3}_e \approx - 0.4\% \left(\frac{s_{13}}{0.16}\right)^2.  
\ee 
(Notice that there is 
a small matter effect on $s_{13}$ which is different in neutrino 
and antineutrino channel.)  

The interference term is linear in $s_{13}$ and positive for the negative 
sign of $s_{13}$:  
\be
\epsilon^{int}_e = - 2.0 \% \left(\frac{s_{13}}{0.16}\right). 
\ee 
It can be estimated as follows. 
The depth of oscillations equals  $\sim \sin 2\theta_{12}^m
\cos 2\theta_{12}^m =  0.47$.  The averaging over $\cos \Theta_{\nu}$ gives 
$\langle R_2 \rangle \sim 0.14$. 
Due to negative effect for antineutrinos the total effect is 
$\langle R_2 \rangle \approx 0.09$. 
Averaging over the energy (which is important here) leads to  further
reduction by about 30 \%. Finally we get $\epsilon^{int}_e = \langle R_2
\rangle r s_{13}  \approx 2\%$ for $s_{13} = -0.16$ in agreement with calculations.  


Summing up all the contribution we find for $s_{13} = - 0.16$: 
$\epsilon_e^{tot}  = 2.5\%$ in agreement with  result 
in fig.~\ref{fig9} (upper panel). For $s_{13} = + 0.16$, the interference term 
changes the sign and we get: 
$\epsilon_e^{tot}  = -1.7\%$. Apparently the curves 
are not symmetric with respect to $N^{osc}_e/N^0_e = 1$ due to the LMA-
and $U_{e3}$- contributions 
which do not change the sign with $s_{13}$. When 
$|s_{13}|$  decreases,  
both $\epsilon^{int}_e$ and $\epsilon^{Ue3}_e$ decrease in absolute value.

In fig.~\ref{fig9} (lower panel) we show the  zenith angle dependences
for larger deviation of  2-3 mixing from maximal value, $s^2_{23}=
0.35$. Now screening is weaker and the LMA oscillations give main
contribution $\epsilon_e^{LMA} = 2.9\%$. 
This leads to the shift of all the histograms to  $N^{osc}_e/N^0_e > 1$.  

Now also the $U_{e3}$ contribution is larger being comparable with the 
interference  contribution: 
\be
\epsilon^{Ue3}_e \approx - 1.3\% \left(\frac{s_{13}}{0.16}\right)^2. 
\ee
In contrast, the interference term has no screening factor 
and its absolute value even slightly decreases in comparison with the 
previous case due to decrease of $\sin^2 2\theta_{23}$: 
\be
\epsilon^{int}_e = - 1.9 \% \left(\frac{s_{13}}{0.16}\right).
\ee
This leads to more complicated dependence 
of the excess on  $s_{13}$. We find that maximum of total excess is 
realized for $s_{13} = - 0.16$:   
$\epsilon_e^{tot} =   3.6\%$. 
For $s_{13} = - 0.08$ we get a very similar value:  
$\epsilon_e^{tot}  =  3.55\%$.

Maximal effect of interference can be estimated in the following way. 
As we have discussed, $R_2^{max} = 1/2$ (which can be achieved at 
$\Delta m^2_{12} = 7.3 \cdot  10^{-5}$ eV$^2$). The averaging over the zenith 
angle gives $\langle R_2 \rangle = R_2^{max}/4 = 0.125$. The antineutrino 
contribution is negative,  
so that $\epsilon^{int}_e < 2 s_{13} (1 - \xi) 0.125 = 0.18 s_{13} < 3\%$.  
Averaging over the energy leads to an additional suppression. 

Using curves which correspond to different sign of  $s_{13}$, 
it is easy to disentangle contribution from the interference term.  
Obviously, 
\be
\epsilon^{int}_e (s_{13}) = \frac{1}{2} 
\left[\epsilon_e^{tot} (s_{13}) - \epsilon_e^{tot} (- s_{13}) \right], 
\label{int-tot}
\ee
and for the two other contributions we get: 
\be 
\epsilon_e^{LMA} = \epsilon_e^{tot} (s_{13} = 0), ~~~~~
\epsilon^{Ue3}_e = \frac{1}{2}
\left[\epsilon^{int}_e (s_{13}) + \epsilon^{int}_e (- s_{13}) \right] -
\epsilon_e^{tot} (s_{13} = 0).  
\label{ue3-tot}
\ee

In Fig.~\ref{fig11} we show dependence of the ratio 
of number of events integrated over the zenith angle,  $N^{osc}_e/N^0_e$, on 
the energy.  According to our analytical 
consideration,  with decrease of energy  the LMA contribution (green
histogram) increases fast, the $U_{e3}$-contribution is unchanged and 
the interference term first, increases but then below $E \sim 0.4$ GeV
starts to decrease. Using relations (\ref{int-tot} - \ref{ue3-tot}) we
find from the Fig.~\ref{fig11} (upper panel) for  $s_{13} = - 0.16$
that in the bins $E = (0.4 - 0.65),~ (0.25 -0.40),~ (0.10 - 0.25)$ GeV:  
$\epsilon_e^{LMA}  = 1.2\%,~~ 2.7\%,~~ 5.9\%$ respectively, 
$\epsilon^{Ue3}_e = -1.1\% = const$,  whereas 
$\epsilon^{int}_e = 1.7 \%,~~ 2.7\%,~~ 2.4\%$. At high energies the
interference term gives main contribution. 
(Notice, however, that for $E > 0.6$ GeV our approximation may not be
precise).

\subsection{CP-violation effects}

Let us consider effects of the CP-violating phase $\delta_{CP}$. 
Notice that if we substitute $I_2$ by its vacuum value 
(eq. (\ref{mue-pr2}) with $\theta_m \rightarrow \theta$),  the
interference term, as is expected,  becomes equal $\epsilon^{int}_e =
1/2 \Delta P$, where $\Delta P$ is the  neutrino-antineutrino CP-
asymmetry.

\begin{figure}[ht]
\centering\leavevmode
\epsfxsize=0.7\hsize
\epsfbox{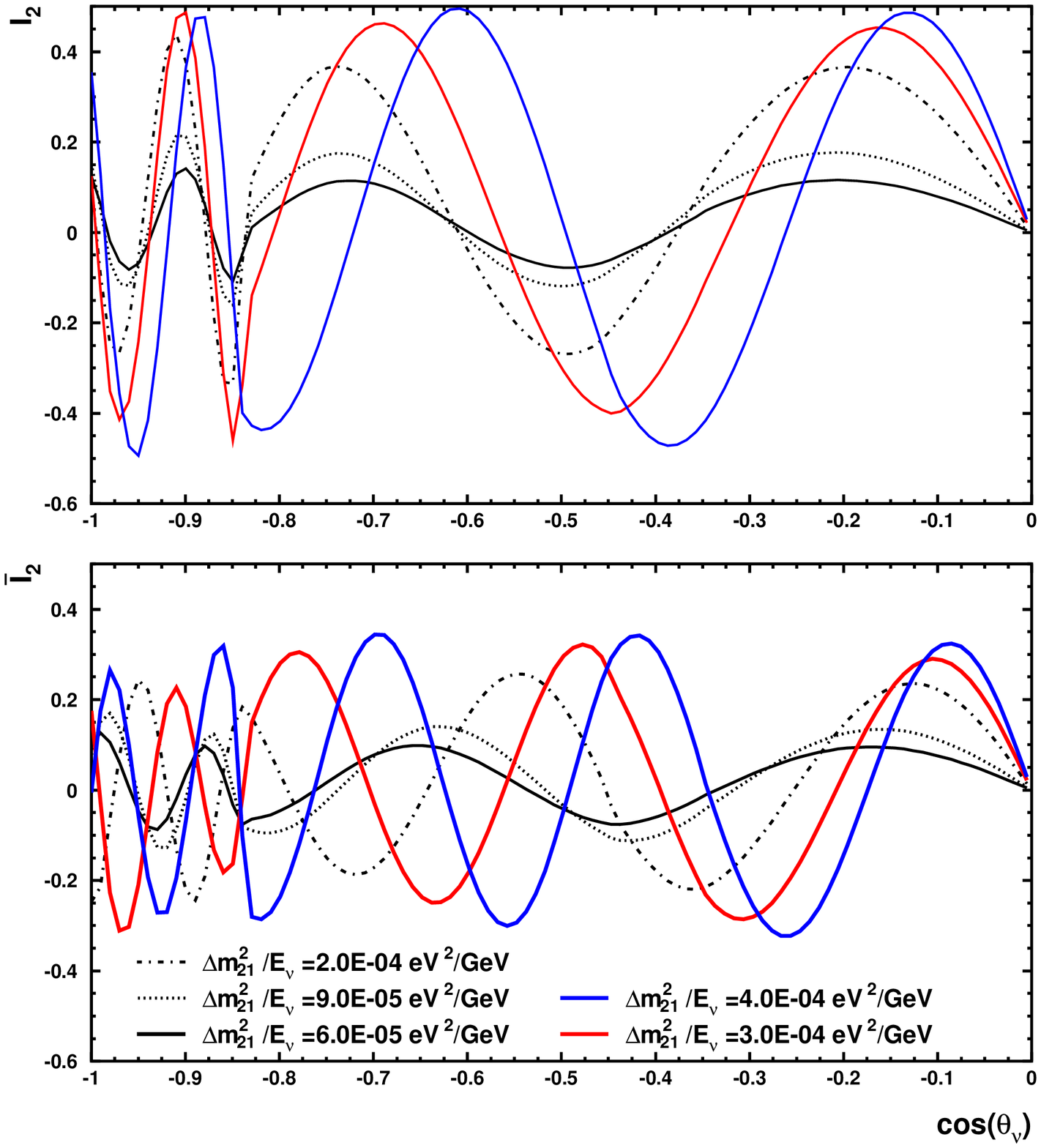}
\caption{
Zenith angle dependences of the imaginary part  of the  interference probability
for neutrinos, $I_2$ (upper panel),  and
for antineutrinos, $\bar{I}_2$  (lower panel),  for different values
of $\Delta m^2_{12}/E$ and $\sin^2 2 \theta_{12} = 0.82$.
}
\label{fig4}
\end{figure}

\begin{figure}[ht]
\centering\leavevmode
\epsfxsize=0.7\hsize
\epsfbox{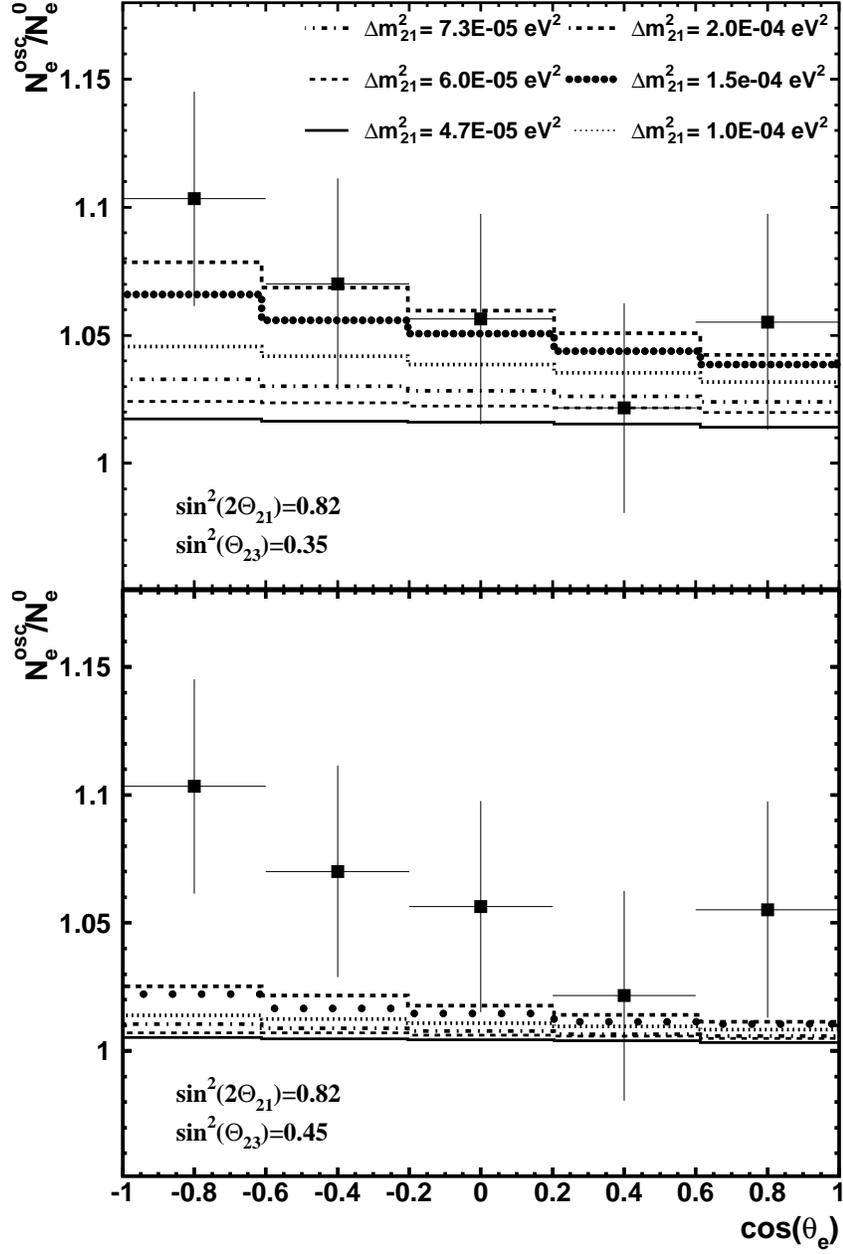}
\caption{
Zenith angle distributions of the $e$-like sub-GeV events for $\sin
\theta_{13} = 0$. The ratio of the numbers of events with and 
without  $\nu_e$-oscillations,  $N_e^{osc}/N_e^0$, 
for different values of $\Delta m^2_{12}$. 
From the upper to lower histograms the corresponding
value of $\Delta m^2_{12}$ decreases.
We take $\sin^2 2 \theta_{12} = 0.82$ and 
$\sin^2 \theta_{23} = 0.35$ (upper panel), 
$\sin^2 \theta_{23} = 0.45$ (lower panel).
Shown are also the Super-Kamiokande  experimental points from
Ref.~\protect\cite{super-data-used}.
}
\label{fig5}
\end{figure}

\begin{figure}[ht]
\centering\leavevmode
\epsfxsize=0.7\hsize
\epsfbox{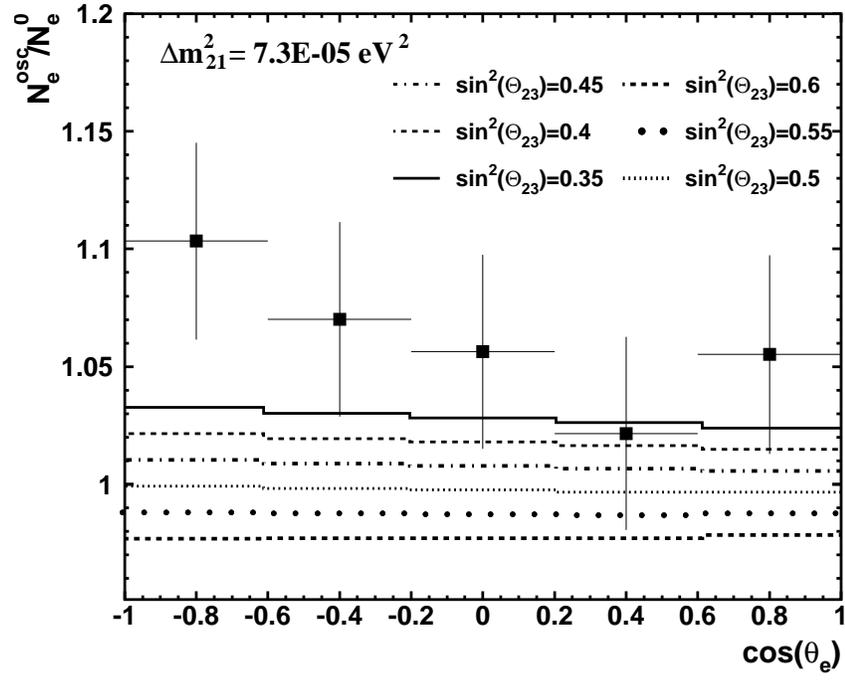}
\caption{
Zenith angle distributions of the $e$-like sub-GeV events for $\sin
\theta_{13} = 0$.  The ratio of numbers of events with and without
$\nu_e$-oscillations,
$N_e^{osc}/N_e^0$, for different values of $\sin^2 \theta_{23}$.   
We take $\sin^2 2 \theta_{12} = 0.82$ and 
$\Delta m^2_{12} = 7.3 \cdot 10^{-5}$~eV$^2$. 
From the upper to lower histograms the corresponding
value of $\sin^2\theta_{23}$ increases.
Shown are also the Super-Kamiokande
experimental points from Ref.~\protect\cite{super-data-used}.
}
\label{fig6}
\end{figure}

\begin{figure}[ht]
\centering\leavevmode
\epsfxsize=0.7\hsize
\epsfbox{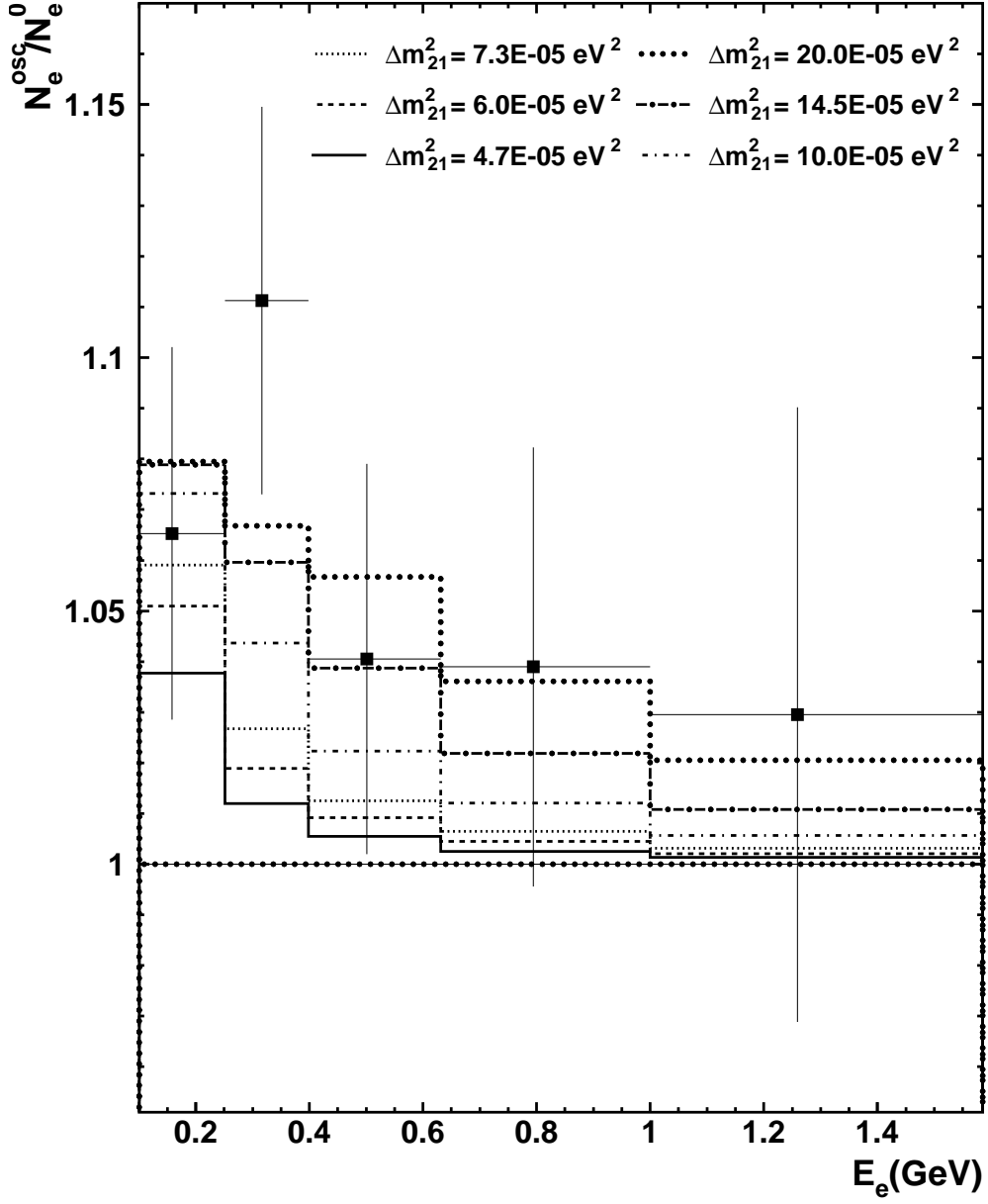}
\caption{
The excess of the $e$- like events integrated over the zenith angles
as a function of energy.  Dependence of the  ratio of numbers of 
the  $e$-like sub-GeV events with and  without $\nu_e$-oscillations,  
$N_e^{osc}/N_e^0$, on the  visible energy for 
different values of $\Delta m^2_{12}$. From the upper to lower
histograms the corresponding value of $\Delta m^2_{12}$ decreases. 
For other parameters we take: 
$\sin^2 2 \theta_{12} = 0.82$, $\sin^2 \theta_{23} = 0.35$ and $\sin
\theta_{13} = 0$. 
Shown are also the Super-Kamiokande  experimental points from
Ref.~\protect\cite{super-data-used}.
}
\label{fig7}
\end{figure}

\begin{figure}[ht]
\centering\leavevmode
\epsfxsize=0.9\hsize
\epsfbox{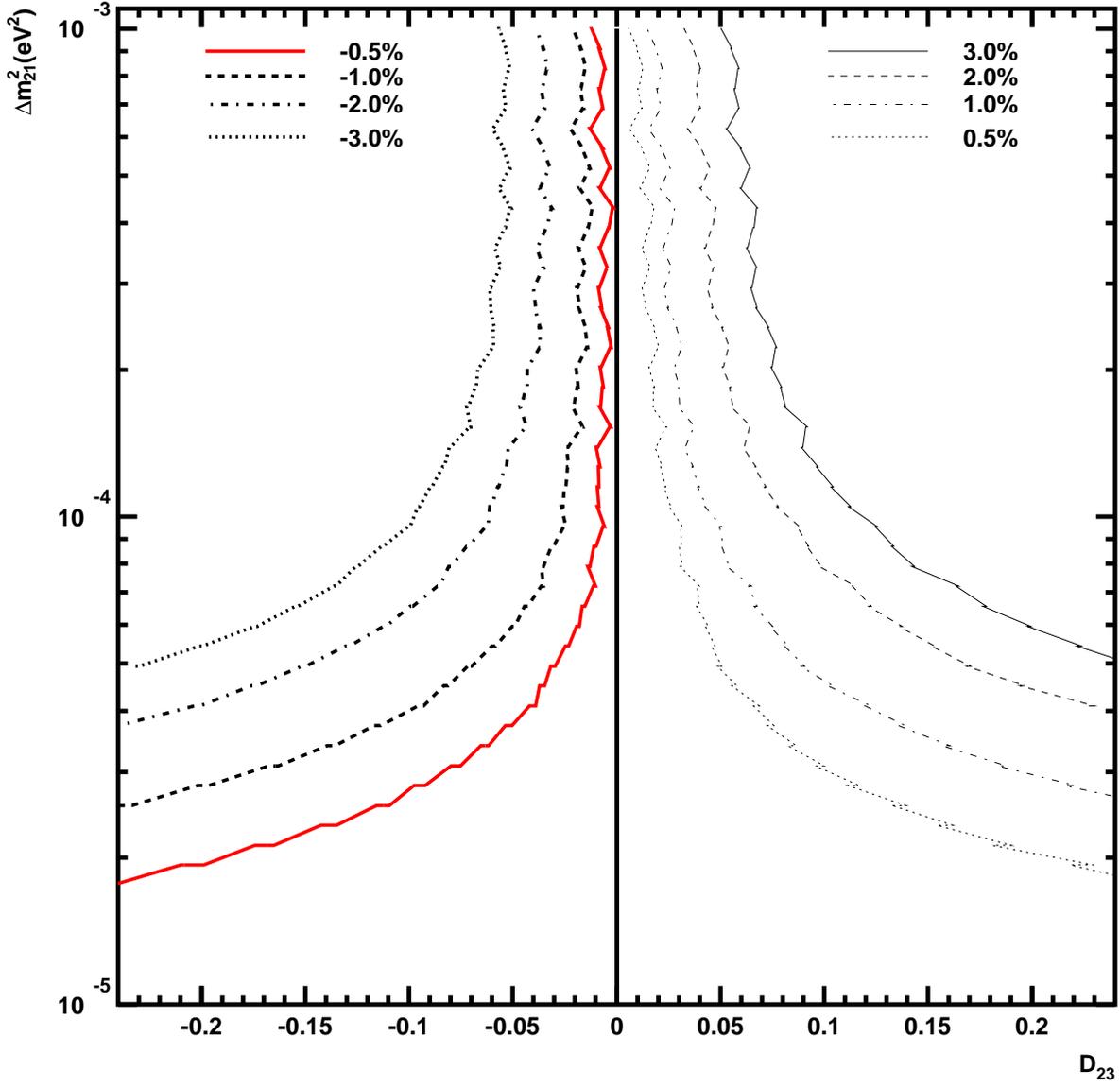}
\caption{ 
The contours of constant relative change of the number  
of the $e$-like events, $N_e^{osc}/N_e^0 - 1$ (in \%), 
in the $\Delta  m^2_{12}$ - 
$D_{23}$ plane. We take $\sin^2 2 \theta_{21} = 0.82$ and
$\sin \theta_{13} = 0$.
}
\label{fig8}
\end{figure}

\begin{figure}[ht]
\centering\leavevmode
\epsfxsize=0.7\hsize
\epsfbox{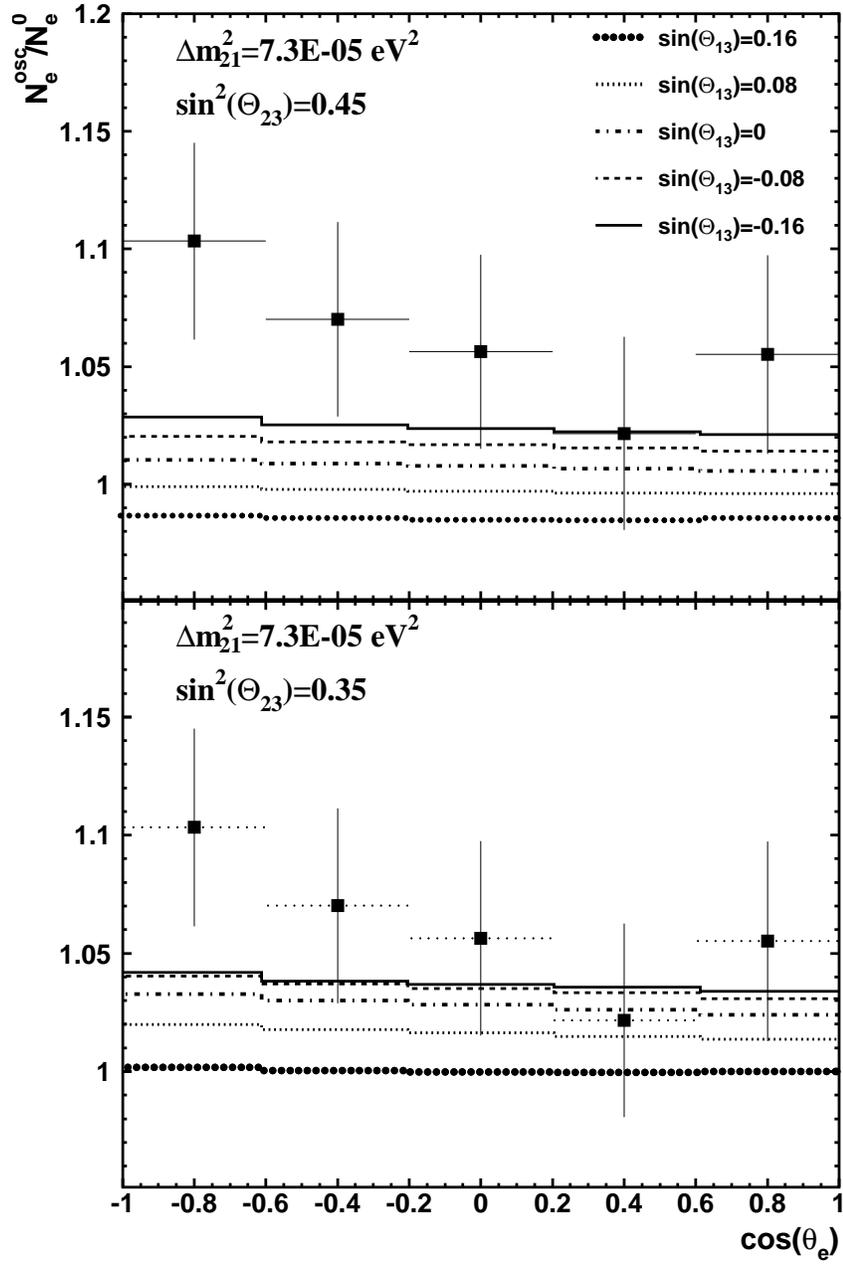}
\caption{
Zenith angle distributions of the sub-GeV $e$-like events for non-zero 1-3 mixing.
The ratio of the  numbers of events with and without $\nu_e$ oscillations 
for different values of $\sin \theta_{13}$.  
We take $\sin^2 2 \theta_{12} = 0.82$, 
$\Delta m^2_{12} = 7.3 \cdot 10^{-5}$~eV$^2$  
and $\sin^2 \theta_{23} = 0.45$ (upper panel),  and 
$\sin^2 \theta_{23} = 0.35$ (lower panel). 
The excess increases with decrease of $\sin \theta_{13}$.
Shown are also the Super-Kamiokande  experimental points from
Ref.~\protect\cite{super-data-used}.
}
\label{fig9}
\end{figure}

\begin{figure}[ht]
\centering\leavevmode
\epsfxsize=0.7\hsize
\epsfbox{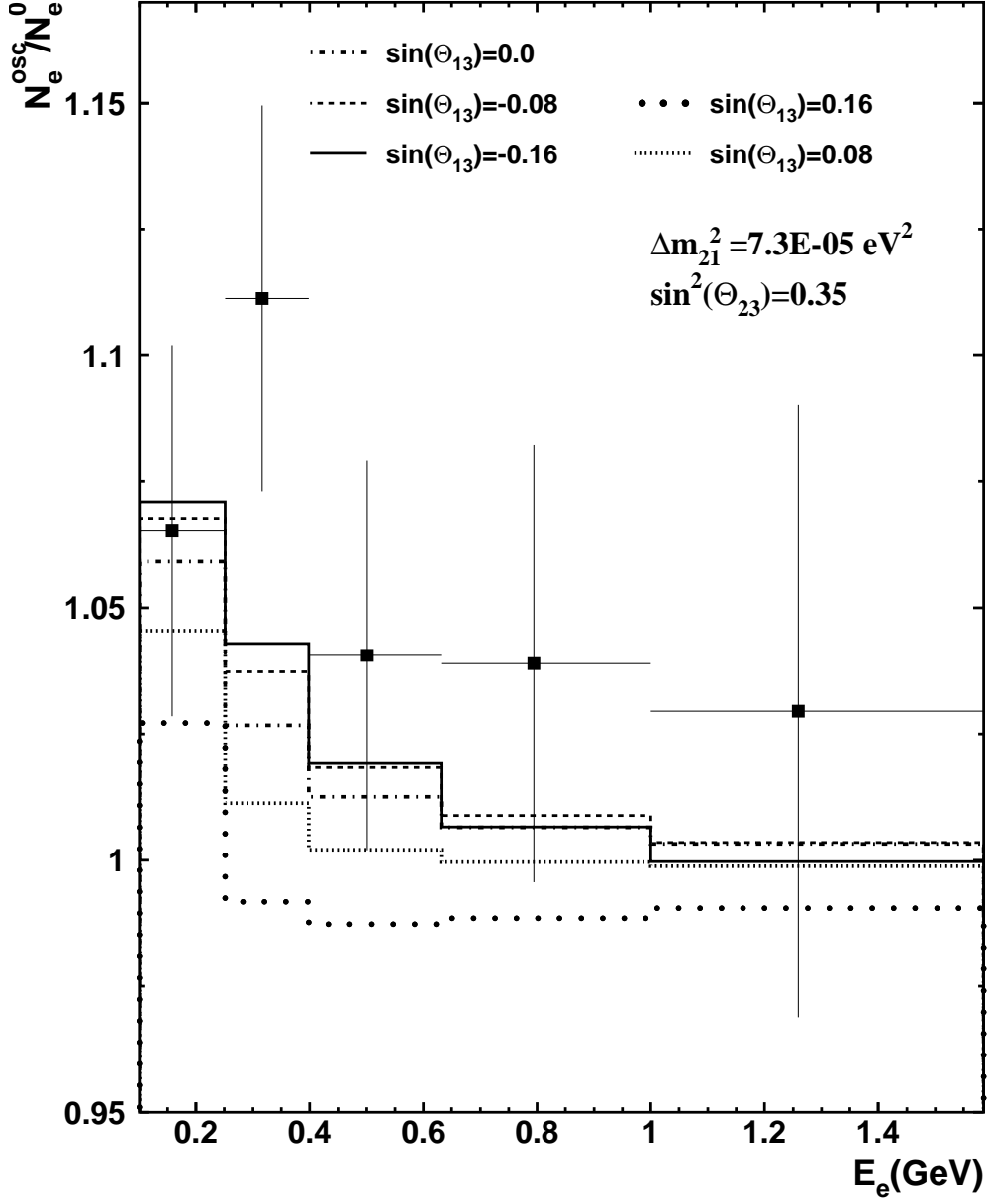}
\caption{
The excess of the $e$- like events integrated over the zenith angles
as a function of energy.  Dependence of the  ratio of number of $e$-like events 
with oscillations and  without oscillations on the  
visible energy for different values of $\sin \theta_{13}$.
For other parameters we take:  $\sin^2 2 \theta_{23} = 0.91$ 
and $\Delta m^2_{12} = 7.3 \cdot 10^{-5}$ eV$^2$.  
The excess increases with decrease of $\sin \theta_{13}$.
Shown are also the Super-Kamiokande  experimental points from
Ref.~\protect\cite{super-data-used}.
}
\label{fig11}
\end{figure}

According to fig.~\ref{fig4}, $I_2$ alternates the sign with change of
the  zenith angle. However, there is no averaging of the effect to
zero for two reasons:

1. For  the mantle trajectories  up to  1.5 - 2 periods 
of oscillations are obtained. In particular,  for 
 $2\cdot 10^{-4}$ eV$^2$/GeV  (fig.~\ref{fig4}) 
which corresponds to the best value of $\Delta m^2_{12}$ and $E= 0.4$ GeV 
there are 1.5 periods. 

2. Due to change of the density for  trajectories with different $\Theta_e$
the curves are not symmetric with respect to $I_2 = 0$. 

For antineutrinos the probability  $\bar{I}_2$ has smaller amplitude
and oscillation length, furthermore,  the curves are nearly symmetric
with respect to $I_2 = 0$. As a results, integration over the zenith
angles leads to strong suppression of the 
averaged value  $\langle I_2 \rangle$. This, as well as  difference of the original 
neutrino and antineutrino fluxes result in  existence of the CP-odd effects,  
even in the sample where neutrino and antineutrino signals  are summed up. 

\begin{figure}[ht]
\centering\leavevmode
\epsfxsize=0.7\hsize
\epsfbox{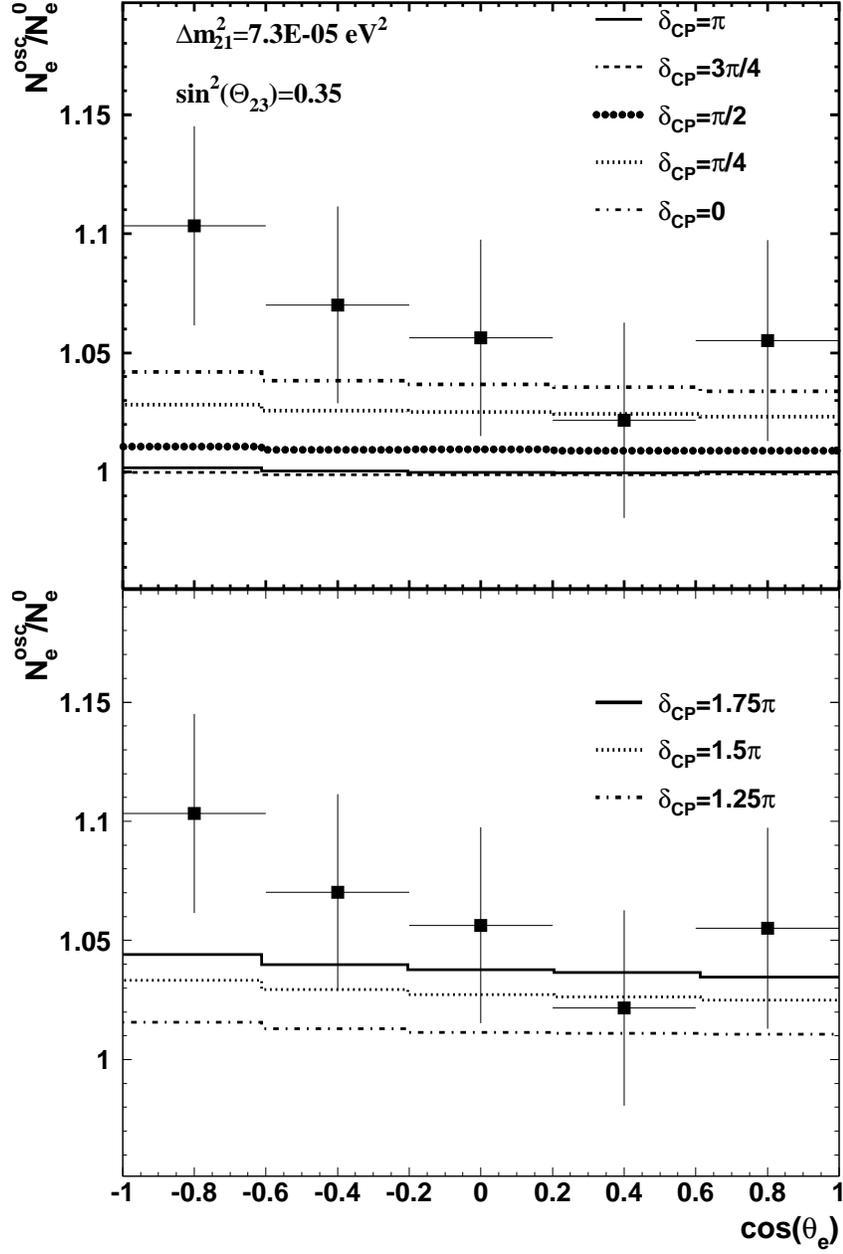}
\caption{
Zenith angle distributions of the sub-GeV $e$-like events for non-zero 1-3 mixing.
The ratio of the  numbers of events with and without $\nu_e$ oscillations
for different values of the CP-violating phase $\delta_{CP}$. 
We take $\sin^2 \theta_{23} = 0.35$, $\sin^2 2 \theta_{12} = 0.82$, 
$\sin \theta_{13} = 0.16$,  
$\Delta m^2_{12} = 7.3 \cdot 10^{-5}$ eV$^2$.
In the upper (lower) panel, the excess decreases (increases) with increase of $\delta_{CP}$.
Shown are also the Super-Kamiokande  experimental points from
Ref.~\protect\cite{super-data-used}.
}
\label{fig12}
\end{figure}

In fig.~\ref{fig12} we show the zenith angle dependence  of the   
oscillation effect for different values of $\delta_{CP}$. 
The analytical expression of this dependence on $\delta_{CP}$ is given
in Eq. (\ref{fl-excessM}). 
Simple estimation of the effect can be obtained as follows.
Using results of fig.~\ref{fig4} we obtain    
after averaging over the zenith angle and 
energies:  $\langle I_2 \rangle = 0.040$ and $\langle \bar{I}_2 \rangle 
\approx 0.018$.  
Then for $s_{13} = - 0.16$ and $s_{23}^2 = 0.35$,   the 
contribution of imaginary part  to the interference term equals: 
$\epsilon_I^{int} \sim  1.1\%$. In the previous section we have found that 
for the same set of parameters $\epsilon_R^{int} \approx 2.0\%$. 
So, the interference effect can be written as 
\be
\epsilon_e^{int} \approx  (2.0 \cos\delta_{CP} -1.1 \sin \delta_{CP}) \%
~~~~(s_{23}^2 = 0.35) . 
\label{epsdelta}
\ee
The relative values of numerical coefficients depend on the $2\nu$ probabilities 
$R_2$ and $I_2$. 
For $\delta_{CP} = -\pi/4$ we find  $\epsilon_e^{int} \approx 2.2 \%$ (no change 
with respect to $\delta_{CP} = 0$). For $\delta_{CP} = \pi/2$: 
$\epsilon_e^{int} = \epsilon_I^{int} \approx  1.1\%$.  Maximal effect of $I_2$ 
is for $\delta_{CP} = - \pi/2$: $\epsilon_e^{int} = \epsilon_I^{int}
\approx 1.1\%$, that is,
 the excess decreases by  $2.8 \%$ in comparison with $\delta_{CP} = 0$ case   
(see fig.~\ref{fig12} upper panel). From fig. \ref{fig9} and relations 
(\ref{int-tot} - \ref{ue3-tot}) we find 
$\epsilon_e^{LMA} + \epsilon^{Ue3}_e \approx 2.2\%$, and consequently, 
\be
\epsilon_e^{tot} = 2.2\% + \epsilon_e^{int}. 
\ee
The latter formula reproduces well the results shown in fig.~\ref{fig12}.

Let us  compare different contributions to the excess of the $e$-like events.
As we have  established the largest contribution can be obtained from
the LMA term. Maximum of $\epsilon^{LMA}_e \sim  6 \%$ is achieved for the
largest possible $\Delta m^2_{12}$, minimal energy and largest
deviation of the 2-3 mixing from maximum. The contribution does not
depend on $s_{13}$.

The interference term  gives maximal contribution, 
$\epsilon^{int}_e \sim 3\%$,  for $\Delta m^2_{12} = 7 \cdot 10^{-5}$
eV$^2$ and maximal possible value $s_{13}$.  It depends very weakly on
the deviation $D_{23}$.

The $U_{e3}$ maximal contribution,  $\epsilon^{Ue3}_e \sim 1-2\%$, is realized 
for the largest possible values of $s_{13}$ and $D_{23}$. It
does not depend on $\Delta m^2_{12}$ in the first approximation.
According to (\ref{fl-excess})  this term has an opposite sign with respect to
the LMA term and therefore partial cancellation with 
the LMA contribution always occurs.

These results allow one to understand that the  contributions of non-zero
$s_{13}$ (interference term and $U_{e3}$) can not further enhance
the excess produced by the LMA term. 
Indeed, for large $\Delta m^2_{12} \sim 2 \cdot 10^{-4}$ eV$^2$  where
$\epsilon^{LMA}_e$ is maximal, the interference term contribution is already
small,
and moreover, the  $U_{e3}$ term is negative compensating substantially the
positive $\epsilon^{int}_e$ contribution. As a result the LMA contribution
can be enhanced by about $1\%$ at most.
For $\Delta m^2_{12} \sim 7 \cdot 10^{-5}$ eV$^2$  where
the interference term is maximal, the LMA contribution is smaller and again
partial cancellation with $U_{e3}$ contribution occurs.

Notice that the cancellation  of the LMA induced excess  can be
stronger than the enhancement since  $\epsilon^{int}_e$ and $\epsilon^{Ue3}_e$ can have both
the same negative sign with respect to $\epsilon^{LMA}_e$.

\section{Measuring $D_{23}$ and $\delta_{CP}$}

In terms of the deviation parameter (\ref{dev}) the excess of the $e$-like
events can be written as
\be
\epsilon_e  =   r \left( \langle P_2 \rangle - 2 \tilde{s}_{13}^2 \right) D_{23} +
\epsilon^{int}_e + (r/2 -1) \left(\langle P_2 \rangle + 2 \tilde{s}_{13}^2 \right).
\label{dev-eps}
\ee
The interference term depends on the deviation very weakly:
$\epsilon^{int}_e \propto \sin 2 \theta_{23} =
\sqrt{1 - 4 D_{23}^2} \approx 1 - 2 D_{23}^2$. Since variations 
of $D_{23}$ in the presently allowed range  (\ref{devb}) 
change  $\epsilon^{int}_e$
by less than 5\% and in the first approximation this dependence can be
neglected. Also the last term in (\ref{dev-eps}) does not exceed 0.5 \%.
Therefore with a good approximation $\epsilon_e$ is a linear function of $D_{23}$. 
We can find coefficients of this function using fig.~\ref{fig9}.

For  $\Delta m^2_{12} \sim 7 \cdot 10^{-5}$ eV$^2$  and zero 1-3 mixing we
get
\be
\epsilon_e^{0}  \approx  (20.5 D_{23}-0.3) \%.
\ee
For $|s_{13}| = 0.16$ maximal (corresponds to
$s_{13} = - 0.16$)  and minimal ($s_{13} =  0.16$)  values of $\epsilon_e$
can be approximated as
\be
\epsilon_e^{max}  = (12.5 D_{23} + 1.8) \%, ~~~~ \epsilon_e^{min} = (12.8 D_{23}-2.1 )\%.
\ee
Since   for the present upper limit on $D_{23}$,
$\epsilon_e^{min}(0.15) < 0$,  any upper experimental bound,  $\epsilon_e^{exp}$, 
will not improve the limit for $D_{23}$ (\ref{devb}).

If the lower bound on $\epsilon_e$ is established,  one can put the
lower bound on $D_{23}$. Using expression for  $\epsilon_e^{max}$
we find that the lower bounds $\epsilon_e^{exp} > 2\%$,  and
$\epsilon_e^{exp} > 3\%$ will give $D_{23} > 0.02$ and $D_{23} >
0.10$ correspondingly.

The bound on $D_{23}$ can be improved if future experiments put 
stronger bound on $s_{13}$. For $|s_{13}| = 0.08$
maximal and minimal  values of $\epsilon_e$ equal 
\be
\epsilon_e^{max}  \approx (17 D_{23} + 0.84) \%, ~~~~ 
\epsilon_e^{min} \approx (18.5 D_{23} - 1.2) \%.
\ee
{}From the expression for $\epsilon_e^{min}$  we find that the present
bound on $D_{23}$  will be improved provided that the upper bound on 
the excess is better than 1.6 \%. 
If  $\epsilon_e^{exp} < 1\%$,  we get  $D_{23} < 0.1$ {\it etc.}.

Using $\epsilon_e^{max}$ we obtain that the lower experimental bound
$\epsilon_e^{exp} >  2\%$ will lead to the lower bound  $D_{23} > 0.07$. 

Suppose very strong bound on  $s_{13}$ is obtained and 
the lower bound $\epsilon_e^{exp} > 0.02$ will be established. Then  
according to fig.~\ref{fig8} the interval
$\Delta m^2_{12} = (7.3 \pm 0.7) \cdot 10^{-5}$ eV$^2$ (10\% error) will 
lead  to the lower bound on deviation from maximal mixing 
$D_{23} >  0.1$.\\ 

Can the CP-violation phase be measured?
As follows from the fig. ~\ref{fig12}, the phase $\delta_{CP}$ does not produce 
any particular zenith angle dependence and the energy dependence (not showed). 
The same effect can be achieved by changing other parameters. 

Let us consider the most favorable case:
$\Delta m^2_{12} \sim 7.3 \cdot 10^{-5}$ eV$^2$ and $|s_{13}| \sim 0.16$.
The total relative change of number of the e-like events can be written as 
\be
\epsilon_e  =
(1.8 \cos \delta_{CP} - 0.8 \sin \delta_{CP}) +   13.6 D_{23} ~\%.
\label{cp-eps}
\ee
Depending on $\delta_{CP}$ the interference term changes in the limits
$- 1.92 \div +1.92$.
The LMA contribution  is restricted by
$|\epsilon^{LMA}_e| \approx 13.6 |D_{23}| \leq 2.04 \%$.
So, the predictions can be in the interval $-3.96 \div +3.96$.

On the other hand, for $\delta_{CP} = 0$, $|\epsilon^{int}_e| = 1.8 \%$. Therefore
for zero CP-violating phase the excess (deficiency) 
can be in  the intervals: $(0.2 \div 3.8) \%$ and $(-3.8 \div -0.2) \%$
which covers practically whole the interval predicted for
non-zero CP-violating phase.
So,  to get any information about $\delta_{CP}$ one needs to improve 
the bound on the deviation $D_{23}$ from independent measurements.
{}From fig.~\ref{fig6} it follows that 
\be
\Delta \epsilon_e =  - 2 \% \cdot \frac{\Delta (\sin^2 \theta_{23})}{0.1}. 
\label{dd23}
\ee
{\it E.g.}, 1\% effect of $\delta_{CP}$ can be produced also by 0.05 change of 
$\sin^2\theta_{23}$: from 0.35 to 0.40. Variations by 
3\% would require the increase  of $\sin^2 \theta_{23}$ from 0.35 to 0.5. 

Apparently, the effect similar to that of   $\delta_{CP}$ can be produced also by small 
variations  of $s_{13}$,   or $\Delta m^2_{12}$. 
Let us consider this degeneracy of parameters.

Using fig.~\ref{fig5} (upper panel) we find that change of the excess by 1\%  
can be achieved by  (35 - 40)\% change of $\Delta m_{12}^2$ near 
the best  fit point,  {\it e.g.}, from 5.2 to $8.7 \cdot 10^{-5}$ eV$^{2}$:  
\be
\Delta \epsilon_e = 0.44 \% \cdot \frac{\Delta (\Delta m_{12}^2)}{10^{-5} {\rm eV}^2} 
\label{dd12}
\ee
(for $\sin^2_{23} = 0.35$). The effect is smaller if 
the  2-3 mixing is closer to maximum. 
$\Delta \epsilon_e = 1$\% would require  rather large change of 
$\Delta m_{12}^2$: from 5 to 15  of $10^{-5}$ eV$^{2}$. 

Further operation of the KamLAND and SNO will allow to 
determine $\Delta m_{12}^2$ with 10\% ambiguity, which 
is transferred in 0.3\% ambiguity in $\epsilon_e$.

The degeneracy of the $\delta_{CP}$ and  $s_{13}$ is more complicated and 
it depends on specific value and sign of $s_{13}$ as well as  $\sin^2\theta_{23}$. 
In particular, for  $\sin^2\theta_{23} = 0.35$ and $s_{13} = - 0.16$ the 
dependence of $\epsilon_e$ on $s_{13}$ is rather weak (fig.~\ref{fig9}): 
A reduction of the excess by 1\% requires  increase of $s_{13}$ from 
- 0.16 to + 0.05. For $\sin^2\theta_{23} = 0.45$, $\Delta \epsilon_e = 1\%$ 
can  be compensated by changes of  $s_{13}$ in the interval 
$- 0.22 \div -0.10$. That is, moderate accuracy of measurements of 
$s_{13}$ could be enough to determine $\delta_{CP}$. Notice, however, that
with decrease of $s_{13}$ the effect of $\delta_{CP}$ decreases. 

Dependence of the oscillation effect on 1-2 mixing is very weak: 
variations of $\tan^2\theta_{12}$ in the interval from 0.30 to 0.52 
produce a change $\Delta \epsilon_e = 0.3\%$. Expected  improvements 
of determination of $\tan^2\theta_{12}$ will further reduce this 
ambiguity.

The main problem is the identification and measurement 
(or restriction) of the oscillation effect in view of  large present uncertainties 
in the original neutrino flux (15 - 20 \%). In principle, if high enough statistics 
will be achieved the oscillation effect 
can be distinguished from the renormalization by its zenith angle 
and  energy dependences. At the same time, further improvements 
in the calculations of the neutrino fluxes are  extremely important. 
Also separate measurements of the neutrino and antineutrino signals will help.

\section{Conclusion}

\noindent
1. After confirmation of the LMA MSW solution of the solar neutrino problem 
it is clear  that the effect of 
$\nu_e$ - oscillations should appear in the atmospheric neutrinos at some 
level  even for zero value of $s_{13}$.

For the allowed values of the oscillation parameters, in particular,  
$\sin^2 2\theta_{23} > 0.91$,  
the LMA oscillations can produce the integrated 
effect (excess or deficit) up to (5 - 6) \% in the sub-GeV sample.  
The effect increases with decrease of energy and 
in the low energy part of the sample it can be as large as 8\%. 
The zenith angle dependence of the effect is rather weak with 
maximum  achieved in the upward-going bins. 
 
The LMA effect is strongly suppressed for exactly maximal 2-3 mixing, so 
searches for the oscillation effect in the sub-GeV sample can be used   
to measure  the deviation of 2 - 3 mixing from maximum. 
Here, however, an ambiguity  appears due to unknown values of $s_{13}$ and $\delta_{CP}$. 
The ambiguity can be reduced if stronger bound on $s_{13}$ will be established from 
independent measurements. 

\noindent
2. The present experimental accuracy is comparable with the maximal expected effect. 
Notice that without additional renormalization of the original
neutrino fluxes, the data show 
some excess of the $e$-like events which can be explained 
(at least partially) by the LMA-oscillations. In fact, 
the data (including weak zenith angle and energy dependences of the excess 
can be perfectly reproduced by  the LMA-contribution corresponding to  the  best 
fit values of parameters and partial (3- 5\%) renormalization of spectrum.

The excess of $e$-like events in the sub-GeV sample and the absence of
the excess in the  multi-GeV range (as it is indicated by the present
data) testify for the deviation of  the 2-3 mixing from maximum.

\noindent
3. Non-zero 1-3 mixing gives an additional contribution to the oscillation 
effect.  
For the sub-GeV sample, it  leads to  interference of the 
the two neutrino amplitudes driven by solar oscillation parameters 
(induced interference). 
The interference term is linear in $s_{13}$ and does not contain 
the screening factor. It dominates if 2-3 mixing is close to maximal.

The interference term can reach 2 - 3\%. The maximal value  (real
part) corresponds to  $\Delta m_{12}^2 \approx 7 \cdot 10^{-5}$ eV$^{2}$ for
the sub-GeV sample. 

The 1-3 mixing leads also to contribution proportional to 
$s_{13}^2$ which does not exceed $\sim 1\%$. 

\noindent
4. The interference term depends on the CP-violating phase 
$\delta_{CP}$. Variations of $\delta_{CP}$ can  change  the 
oscillation effect by $|\Delta \epsilon| = 3$\%.

\noindent
5. The relative effects of $\delta_{CP}$ is  enhanced for the sample
induced by neutrinos  or antineutrinos, that is, when the sign of 
the electric charge of electron is identified. 

\noindent
6. Various contributions  to the oscillation effect have slightly 
different zenith angle and also  energy dependences. 
This, in principle,  can be used to separate them. 
In particular, the LMA-contribution increases with decrease of the energy, 
the interference term first increases  and then starts to decrease. 

The zenith angle dependence is very weak for low energy bins. 

\noindent
7. There is strong ``degeneracy'' of parameters once total excess is 
measured only. The same integral oscillation effect can be produced for 
different values of $\sin^2\theta_{23}$, $\Delta m^2_{12}$, 
$s_{13}$ and $\delta_{CP}$.

\noindent
8. In principle, future high statistics studies of the atmospheric 
neutrinos  will allow to measure the neutrino oscillation parameters. 
For this the accuracy of measurements of the oscillation effect 
should be about 1\% or better. Also  a way should be found 
to distinguish the oscillation effect  from the 
effect of the neutrino flux normalization. 
The problem of  degeneracy of parameters should be resolved. 
There is a good chance to measure  $\Delta m^2_{12}$ 
with high enough accuracy, so that the corresponding uncertainty will be eliminated. 
It will be very difficult to resolve ambiguity related to 
of $\sin^2\theta_{23}$, $s_{13}$ and $\delta_{CP}$. 
If future  ({\it e.g.}, reactor) experiments put stronger 
bound on $s_{13}$, the ambiguity related to $s_{13}$ and $\delta_{CP}$ can be 
substantially 
reduced. This will allow to use the  atmospheric data to restrict a 
deviation  of the 2-3 mixing from the  maximal one.

\noindent
9. With present knowledge of the oscillation parameters, 
one can expect the effect of $\nu_e$ oscillations at the level of  
existing  experimental error bars and uncertainties in the normalization 
of fluxes. The effect of the LMA oscillations  should be taken into
account in the analysis of the  atmospheric neutrino data.

\begin{acknowledgments} 
This work was supported by Funda\c{c}\~ao de Amparo
 \`a Pesquisa do Estado de S\~ao Paulo (FAPESP), Conselho
Nacional de Ci\^encia e Tecnologia (CNPq), DGICYT under grant
PB95-1077 and by the TMR network grant ERBFMRXCT960090 of the European Union. 
O.L.G.P. thanks M.C. Gonzalez-Garcia and H. Nunokawa for
the atmospheric neutrino code and T. Stanev for the table of
atmospheric neutrino fluxes. O.L.G.P. is  grateful for the hospitality of The Abdus Salam
International Centre for Theoretical Physics, where this work has been 
completed. 
\end{acknowledgments}

\section*{Appendix}

Let us evaluate the effect of the interference between the solar 
and the atmospheric frequencies we have neglected in our consideration. 
This interference  gives additional terms to the probabilities 
$P_{\mu e}$ (\ref{mue-pr}) and $P_{ee}$ (\ref{ee-pr}): 
\be
\Delta P_{\mu e} = 
- 2 \tilde{s}_{13}^2 \tilde{c}_{13}^2 s_{23}^2 K_{ee} + 
2 \tilde{s}_{13} \tilde{c}_{13}^2 s_{23} c_{23} K_{\mu e}, 
\label{del-mu}
\ee
\be
\Delta P_{ee} = 2 \tilde{s}_{13}^2 \tilde{c}_{13}^2 K_{ee},
\label{del-ee}
\ee
where 
\be 
K_{ee} \equiv 
Re \left[\tilde{A}_{ee}^* \tilde{A}_{\tau \tau} \right],~~~
K_{\mu e} = Re \left[\tilde{A}_{\mu e}^* e^{i\delta_{CP}} \tilde{A}_{\tau
\tau} \right],~~~
\label{kk}
\ee
Notice that $K_{\mu e}$ depends on the CP-violating phase. 
Inserting (\ref{del-mu}) and (\ref{del-ee}) in (\ref{fluxe-osc}) we get 
the corresponding corrections to the relative change of number of the
$e$-like events:  
\be
\Delta \epsilon_e = 
2 \tilde{s}_{13}^2 \tilde{c}_{13}^2 (1 - r s_{23}^2) \langle K_{ee} \rangle +
2 r \tilde{s}_{13} \tilde{c}_{13}^2 s_{23} c_{23} \langle K_{\mu e} \rangle . 
\label{delta-e}
\ee
Here $\langle ... \rangle$ denotes the averaging over the energy and
the zenith angle. 

Let us evaluate two terms in this equation in order. 

1). The first term is proportional to two small factors: 
$s_{13}^2 D_{23} < 0.015$.  
For trajectories with $\cos \Theta > 0$ taking $\tilde A_{ee} \approx
1$ we estimate:  $\langle K_{ee} \rangle < \langle \cos \Phi_3 \rangle$. 
Since for typical energy 0.4 GeV, the oscillation length in vacuum
is $l_{13} \sim 500$ km, 
and the phase equals $\Phi_3 \sim 2\pi$. Therefore 
averaging over the zenith angle and the energy lead to 
strong suppression: $\langle K_{ee} \rangle \sim 0.2$. As a result, 
the whole term is smaller than 0.3\%.

2). In the second term of (\ref{delta-e}) $\langle K_{\mu e} \rangle$ can be estimated in
the following way.  In vacuum: 
\be
\tilde A_{\mu e} = - s_{12} c_{12} \left( 1 - e^{-i\phi_2} \right), ~~~ 
\phi_2 = \frac{\Delta m_{12}^2 L}{2 E} . 
\ee
For trajectories with $\cos \Theta > 0$ the phase driven by the solar mass split 
is small: $\phi_2 < 0.2$, so that 
\be
\langle K_{\mu e} \rangle \approx - s_{12} c_{12} 
\langle \phi_2 \sin(\phi_3 + \delta_{CP})\rangle
= - s_{12} c_{12} \langle \frac{2 \pi d}{l_{12}} \frac{1}{\cos\Theta}
\sin(\phi_3 + \delta_{CP})\rangle ,
\label{kkk}
\ee
where $d \sim 20$ km is the depth of the atmosphere. 
For  $E = 0.4$ GeV the oscillation length equals $l_{12} = 1.4 \cdot 10^{4}$
km. The averaging over the zenith angle gives $\langle 1/\cos \Theta
\rangle = 3.2$. Then taking $\sin(\phi_3 + \delta_{CP}) = 1$ we obtain from (\ref{kkk}) 
\be
|\langle K_{\mu e} \rangle| <   3 \cdot 10^{-2} s_{12} c_{12}, 
\ee
and consequently, for  $s_{13} \leq 0.16$ the contribution of the second term to 
$\Delta \epsilon_e < 0.5 \%$. 
Averaging over the energy leads to further suppression of this contribution.


\newpage


\end{document}